\documentclass[useAMS,usegraphicx,referee]{mn2e}
\usepackage{lscape}

\title[Nuclear activity in ULIRGs]{The role of nuclear activity as the power source 
of ultraluminous infrared galaxies}
\author[E. Nardini et al.]{E.~Nardini,$^1$\thanks{E-mail: nardini@arcetri.astro.it}
G.~Risaliti,$^{2,3}$ Y.~Watabe,$^2$ M.~Salvati,$^2$ E.~Sani$^1$\\
$^1$ Dipartimento di Fisica e Astronomia - Sezione di Astronomia, Universit\`a di Firenze, 
L.go E. Fermi 2, 50125 Firenze, Italy\\
$^2$ INAF - Osservatorio Astrofisico di Arcetri, L.go E. Fermi 5, 50125 Firenze, Italy\\
$^3$ Harvard-Smithsonian Center for Astrophysics, 60 Garden St. Cambridge, MA 02138 USA}

\begin{document}

\date{Released Xxxx Xxxxx XX}

\pagerange{\pageref{firstpage}--\pageref{lastpage}} \pubyear{2010}

\maketitle

\label{firstpage}

\begin{abstract}
We present the results of a 5--8~$\mu$m spectral analysis performed on the largest sample of local 
ultraluminous infrared galaxies (ULIRGs) selected so far, consisting of 164 objects up to a redshift of 
$\sim$0.35. The unprecedented sensitivity of the Infrared Spectrograph onboard \textit{Spitzer} allowed 
us to develop an effective diagnostic method to quantify the active galactic nucleus (AGN) and starburst 
(SB) contribution to this class of objects. The large AGN over SB brightness ratio at 5--8~$\mu$m and 
the sharp difference between the spectral properties of AGN and SB galaxies in this wavelength 
range make it possible to detect even faint or obscured nuclear activity, and disentangle its emission 
from that of star formation. By defining a simple model we are also able to estimate the intrinsic bolometric 
corrections for both the AGN and SB components, and obtain the relative AGN/SB contribution to the total 
luminosity of each source. Our main results are the following: \\
1) The AGN detection rate among local ULIRGs amounts up to 70 per cent, with 113/164 
convincing detections within our sample, while the global AGN/SB power balance is $\sim$1/3. \\
2) A general agreement is found with optical classification; however, among the objects with no 
spectral signatures of nuclear activity, our IR diagnostics find a subclass of  \textit{elusive}, highly obscured AGN. \\
3) We analyse the correlation between nuclear activity and IR luminosity, 
recovering the well-known trend of growing AGN significance as a function of the overall energy output of the 
system: the sources exclusively powered by star formation are mainly found at 
$L_\mathit{IR}<10^{12.3} L_\odot$, while the average AGN contribution rises from $\sim$10 to $\sim$60 per 
cent across the ULIRG luminosity range. \\
4) From a morphological point of view, we confirm that the AGN content is larger in compact systems, but the 
link between activity and evolutionary stage is rather loose. \\
5) By analysing a control sample of IR-luminous galaxies around $z \sim 1$, we find evidence for only minor 
changes with redshift of the large-scale spectral properties of the AGN and SB components. This underlines the 
potential of our method as a straightforward and quantitative AGN/SB diagnostic tool for ULIRG-like systems 
at high redshift as well, and hints to possible photometric variants for fainter sources. 

\end{abstract}


\section{Introduction}

Ultraluminous infrared galaxies (ULIRGs, $L_\mathit{IR} \sim L_\mathit{bol} > 10^{12} L_\odot$) 
emit the bulk of their energy at 8--1000~$\mu$m, and are the most luminous among the 
local sources (Sanders \& Mirabel 1996; Lonsdale, Farrah \& Smith 2006). Moreover, their 
high-redshift counterparts represent a key component of the early Universe (Blain et al. 
2002; Caputi et al. 2007). Providing a comprehensive picture of the neighbouring population 
is therefore a task whose implications are manifold and far-reaching. In the last years a 
consensus view has rapidly grown about ULIRGs, thanks to the multiwavelength approach to 
their study. Most of these systems are the result of a major merger (e.g. Kim, Veilleux 
\& Sanders 2002; Dasyra et al. 2006a), that may eventually lead to the formation of elliptical 
galaxies with moderate mass (Dasyra et al. 2006b). According to numerical simulations, the tidal 
interaction between the progenitors drives an inflow of gaseous material that triggers and 
feeds both intense star formation and black hole accretion (Mihos \& Hernquist 1996; 
Springel, Di Matteo \& Hernquist 2005). The primary radiation field from the starburst (SB) 
and active galactic nucleus (AGN) components is reprocessed by the surrounding dust, giving 
rise to the huge infrared (IR) emission of ULIRGs. Anyway, the great opacity of the nuclear 
environment hinders a clear identification of the underlying power source, and the detection 
of faint or highly obscured AGN components inside ULIRGs has always been a major challenge. 
The most effective way to address this matter is to search for relic signatures of the buried 
engine within the dust emission itself. The advent of the Infrared Spectrograph (IRS; Houck 
et al. 2004) onboard the \textit{Spitzer Space Telescope} (Werner et al. 2004) has opened a 
new era in the study of ULIRGs, providing access to a wealth of diagnostic tools at 
$\sim$5--35~$\mu$m, such as the [Ne~\textsc{v}] $\lambda$14.32 high-ionization line, the 
9.7~$\mu$m silicate absorption, the complex of polycyclic aromatic hydrocarbon (PAH) features 
at 6.2--11.3~$\mu$m, and the 6, 15 and 30~$\mu$m continuum colours (e.g. Farrah et al. 2007; 
Spoon et al. 2007; Veilleux et al. 2009a). In the end, the existence of a sizable 
population of AGN which are elusive not only in the optical (Maiolino et al. 2003) but also 
at mid-IR wavelengths has been safely ruled out. \\
Thanks to the high accuracy of the \textit{Spitzer}-IRS observations, we have developed 
an AGN/SB decomposition method based on 5--8~$\mu$m rest-frame spectroscopy, that has 
been successfully tested on a representative sample of local ULIRGs (Nardini et al. 2008, 2009; 
hereafter Paper~I and Paper~II, respectively). At 5--8~$\mu$m an accurate determination of the 
AGN and SB components can be obtained by means of spectral templates: the large difference 
between the average spectral properties of SB galaxies and AGN, along with the little dispersion 
within the separate classes (e.g. Brandl et al. 2006; Netzer et al. 2007), makes this wavelength 
range very favourable to solid ULIRG diagnostics. We are therefore able to unveil even faint or 
obscured nuclear activity, and to assess its contribution to the luminosity of each source. 
In the present work we investigate the properties of the AGN 
population within the largest sample of local ULIRGs studied so far, consisting of 164 objects and 
extending to redshifts up to $z \simeq 0.35$. This allows us to better constrain the correlation 
between nuclear activity and extreme IR emission. In fact, the incidence of black hole accretion on the 
luminosity of ULIRGs is known to increase with the total energy output of the system, from both optical 
classification (Veilleux, Kim \& Sanders 1999) and early mid-IR spectroscopy (e.g. Tran et al. 2001). 
Such a trend is indeed suggested to involve also the IR systems at lower luminosities, as recently 
confirmed by many \textit{Spitzer}-based studies (e.g. Imanishi 2009; Valiante et al. 2009). We also 
take into account a compilation of IR-luminous galaxies at $z \sim 1$, extracted from different samples 
in the literature, in order to test our method against the possible evolution with redshift of the AGN/SB 
templates and bolometric corrections (i.e. the ratios between the 5--8~$\mu$m and the 8--1000~$\mu$m 
luminosities). This paper is arranged as follows: in the first part (Sections~2--3) we describe the 
selection of our local sample and the data reduction. In Section~4 we briefly summarize 
the main steps of our diagnostic method, introducing the model and computing the AGN and 
SB bolometric corrections from which the relative AGN/SB contribution can be derived. 
We then discuss how our results improve the optical classification and fit into the question 
of the growing AGN contribution with IR luminosity (Section~5). The application of our method to 
high-redshift sources and possible, alternate variants are dealt with in Section~6, while the 
conclusions are drawn in Section~7. Throughout this paper we have made use of the concordance 
cosmology from the \textit{Wilkinson Microwave Anisotropy Probe} (\textit{WMAP}) sky survey, with 
$H_0=70.5$~km~s$^{-1}$~Mpc$^{-1}$, $\Omega_m=0.27$ and $\Omega_\Lambda=0.73$ (Hinshaw et al. 2009).

\section{ULIRG sample}

\begin{figure}
\includegraphics[width=8.5cm]{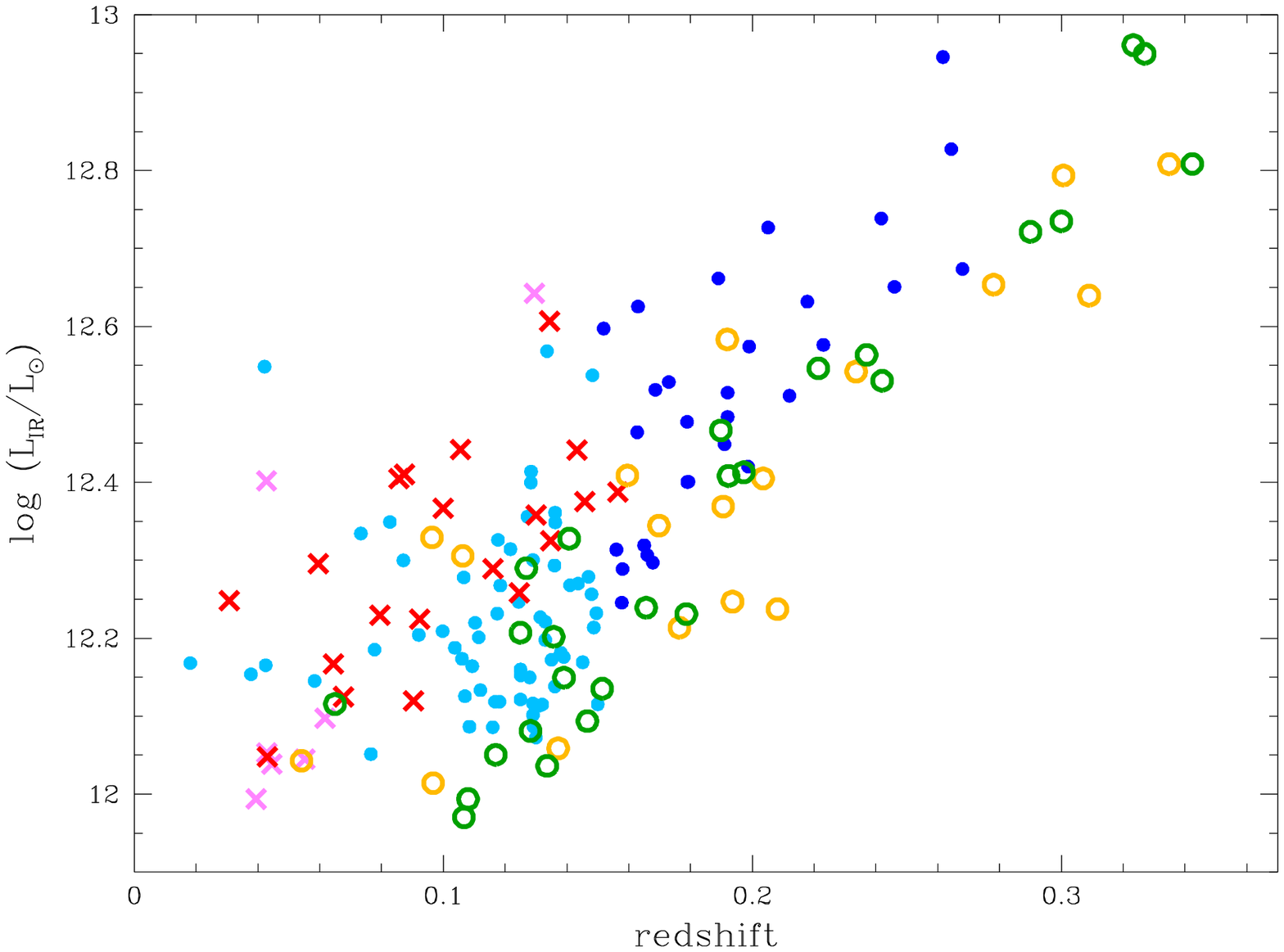}
\caption{The distribution in redshift and IR luminosity of the 164 ULIRGs in our sample. 
The colour and symbol code is attributed as follows, according to the main selections: filled circles 
for sources drawn from the 1~Jy sample (Kim \& Sanders 1998), light blue for those already analysed in 
our previous works, deep blue for additional ones; crosses for sources belonging to the 2~Jy sample 
(Strauss et al. 1992), violet and red, respectively, with the same distinction as above; open circles, green 
for sources in the QDOT catalogue (Lawrence et al. 1999), orange for other sources.}
\label{zl}
\end{figure}

In the wake of the indications from many previous studies, in Paper~II we already looked for the possibility 
of a larger AGN incidence at higher luminosities, and found evidence of an increasing trend. Our preliminary 
results were anyway incomplete due to the limited statistics. We note that the existence of such a correlation is a 
cornerstone in the present knowledge of ULIRGs, but remains a rather \textit{qualitative} statement that has never 
received a \textit{quantitative} treatment. In particular, in order to achieve a comprehensive description of the entire 
ULIRG luminosity range, especially at the higher end, one has to abandon the widely adopted flux threshold of 
1~Jy at 60~$\mu$m (Kim \& Sanders 1998) and to allow for fainter objects. 
In the light of these considerations, the optimal candidates for this study have to be 
found within the \textit{IRAS} PSC$z$ survey (Saunders et al. 2000), which covers the 84 per cent of the 
sky down to a flux density of $\sim$0.6~Jy at 60~$\mu$m. The PSC$z$ catalog 
contains more than a thousand ULIRGs, almost two hundreds of which have been observed by \textit{Spitzer}. 
Among the latter, no archival data are available at present for $\sim$15 sources of the 1~Jy sample, and a handful 
of the observed objects were not detected at 5--8~$\mu$m. This seems to be due to an offset in the target 
pointing and/or to an insufficient exposure time (see the Appendix for more detail). Within the remaining entry list, 
we have applied an additional filter related to the quality of the available measure of the IR luminosity, as explained 
later on. We have also decided to drop 3C~273, since its IR luminosity lies in the ULIRG range only as a consequence 
of its quasar nature, and represents a minor fraction of the bolometric emission. Our final sample consists of 164 sources, 
of which 70 have been already analysed in our previous works.\footnote{NGC~6240 is the only missing source with 
respect to Paper~II, because of its low IR luminosity.} The highest redshift is $z=0.342$, and the luminosity range 
$10^{12} < L_\mathit{IR}/L_\odot < 10^{13}$ is adequately probed in all its extent (see Fig.~\ref{zl}). We stress that 
this sample is fully representative of the ULIRG population in the local Universe. The flux limitation at 60~$\mu$m, in fact, 
translates into a fairly unbiased selection with respect to the nature of the energy source, since $f_{60}$ is a good proxy 
of the cold dust component. Moreover, the core subset (about a hundred objects) is represented by the 1~Jy ULIRGs, and 
forms an almost neutral sample by itself. The extra sources are generally part of extensive observational programmes, 
and are not scheduled by virtue of peculiar properties or unusual activity. No bias towards or against AGN activity is therefore 
expected to undermine our conclusions. In order to study the correlation between AGN contribution and ULIRG luminosity, 
however, both quantities have to be measured with \textit{high} precision. In particular, establishing a reliable and uniform 
method for computing the IR luminosity of the sources is a critical point. 

\subsection{Luminosity measure: uncertainty and \textit{K}~correction effects}

Concerning local ULIRGs, the recurrent and soundest prescription to measure IR luminosities is based on the broad-band 
flux equation of Sanders \& Mirabel (1996):
\begin{equation}
F_\mathit{IR}=1.8 \times 10^{-11}(13.48 f_{12}+5.16 f_{25}+2.58 f_{60}+ f_{100}),
\label{e1}
\end{equation}
where the 8--1000~$\mu$m energy output $F_\mathit{IR}$ (in erg s$^{-1}$ cm$^{-2}$) is obtained as a function of 
the \textit{IRAS} flux densities $f_{12}$, $f_{25}$, $f_{60}$ and $f_{100}$ (in Jy). A consequence of having expanded 
the redshift range is the possible rise of hidden \textit{K}~corrections connected with equation~(\ref{e1}). This problem 
has been examined quantitatively by approximating the expected thermal emission of ULIRGs in the far-IR through a 
modified blackbody function, $S_\nu \propto \nu^{\beta} B_\nu(T_d)$. Disregarding the correlation between the 
emissivity index $\beta$ and the dust temperature $T_d$ (e.g. Yang \& Phillips 2007), we have assumed 
$\beta=1.5$ (as in Yang et al. 2007) and $T_d=35$ and 50~K respectively, to reproduce the spectral energy 
distributions (SEDs) of both SB-like and AGN-like sources.\footnote{The chosen values represent the observed 
limits of the dust temperature range in this greybody scenario, provided that the suspected radio-loud objects are 
excluded (Yang et al. 2007).} We have tested the behaviour of equation~(\ref{e1}) as a function of redshift by 
applying it to the template SEDs; since the $f_{12}$ and $f_{25}$ terms do not contribute much, we focus 
first on the dependence from $f_{60}$ and $f_{100}$. According to the adopted prescription, these far-IR 
terms encompass the bulk (usually more than 70--80 per cent) of the total IR flux of a ULIRG. It turns out that 
$F_\mathit{IR}$ is overestimated by 12 per cent with respect to the rest-frame case for a source at $z=0.3$ 
whose SED is \textit{hot} (i.e. $T_d=50$~K), while it is underestimated by 17 per cent if the SED is \textit{cold} 
(i.e. $T_d=35$~K). The latter effect is 
due to the inadequate sampling of the far-IR emission peak. We now take into account the role of $f_{12}$ 
and $f_{25}$. For most of the sources under study, the mid-IR \textit{IRAS} flux densities are 
expected to be upper limits only. In our previous works we have assumed $f_{12}$ and $f_{25}$ to be 
one half of the available upper limits: at $z \sim 0.1$ this is a good approximation of the actual flux densities, 
as proved by the \textit{Spitzer} spectra. At higher redshifts these upper limits become less constraining. 
The convention we have introduced is anyway acceptable: in fact, a large AGN content involves a hot 
dust component, hence we still have a good evaluation of the true mid-IR emission. On the other hand, 
the likely overestimation of the true $f_{12}$ and $f_{25}$ in SB-dominated sources partially compensates 
for the above-mentioned failure in detecting the peak of a \textit{cold} SED. In conclusion, the possible 
scatter due to \textit{K}~correction effects on equation~(\ref{e1}) are safely within 20 per cent, i.e. $<0.1$~dex. 
We will see in the following that the typical uncertainty of our analysis, when considering the individual objects, 
can be as large as $\sim$0.3~dex. This somewhat random effect is then negligible, and equation~(\ref{e1}) 
with the above prescription for upper limits 
ensures a homogeneous evaluation of IR luminosity for the local ULIRG population. In some cases, 
anyway, the required accuracy can not be achieved, as anticipated before. This occurs when either the 
contribution to $F_\mathit{IR}$ from the \textit{dummy} $f_{12}$ and $f_{25}$ terms is larger than 
$\sim$40 per cent, or the measure of $f_{100}$ is not well-defined. These problems involve $\sim$10 objects 
that have not been included in the sample, leading to our final list of 164 sources.

\section{Data reduction}

The spectroscopic observations were carried out with the \textit{Spitzer}-IRS low-resolution 
modules, within various programmes dedicated to ULIRGs and merging systems; the basic details 
can be found in Table~\ref{t1}. Our reduction strategy has been fully discussed in Paper~II. 
We are dealing with fairly bright objects, whose emission can be easily distinguished from the 
background (mostly due to zodiacal light): we have therefore made use of the coadded images, 
i.e. the final data products provided by the \textit{Spitzer Science Center} after the default 
processing of the individual snapshots.\footnote{The pipeline versions involved in this work 
range from 13.0 to 18.7.} As a single observation in staring mode consists of two different 
exposures, the companion images in the nodding cycle have been subtracted from each other to 
remove the background emission. The spectral extraction has been performed following the 
standard procedure for point-like sources within the software \texttt{SPICE}, for both the 
\textit{positive} and \textit{negative} track. The companion spectra have then been averaged. 
For some fainter objects, a slightly different technique has been adopted to improve the 
signal-to-noise ratio. We have used directly the individual snapshots, performing the 
background subtraction at this stage and replacing the co-addition of the bidimensional 
images with the final average of the corresponding spectra. \\
The major concern in the subsequent phase is the accuracy of the flux calibration. 
As detailed in the following, a good precision is necessary to achieve a reliable estimate of 
the relative AGN/SB contribution to the bolometric luminosity. In the earlier versions of the 
processing pipeline involved here, the uncertainty on the absolute flux calibration was claimed 
to be less than 20 per cent (IRS Data Handbook). Even in this case, our results are not significantly 
affected. We note that any attempt at rescaling our spectra in order to match the \textit{IRAS} 
photometry is excluded, since only upper limits are available at 12 and 25~$\mu$m for 
most of the sources in the sample. We have anyway checked that the \textit{IRAS} constraints 
lie above the \textit{Spitzer} measurements, especially in the presence of a discrepancy among the normalization of the 
different spectral orders. In fact, even if interested in the 
5--8~$\mu$m range only, we have performed the spectral extraction of both the Short-Low (SL) 
and Long-Low (LL) orders and the entire $\sim$5--35~$\mu$m \textit{Spitzer}-IRS spectra are 
available. With a few exceptions, the connection among the orders is smooth or within 
the expected scatter ($\sim$5 per cent). In a limited number of nearby sources the SL1 flux 
(slit width of 3.7 arcsec) is smaller than the LL2 flux (slit width of 10.5 arcsec) by 30--70 
per cent. This is interpreted as an aperture loss, and the SL1 spectra have been scaled up. 
After the comparison with the \textit{IRAS} constraints and the correction of the latter effect, 
we conclude that the calibration of all the 5--8~$\mu$m spectra in our sample is accurate within 
20 per cent at worst. \\

\section{Data analysis}

\begin{figure}
\includegraphics[width=8.5cm]{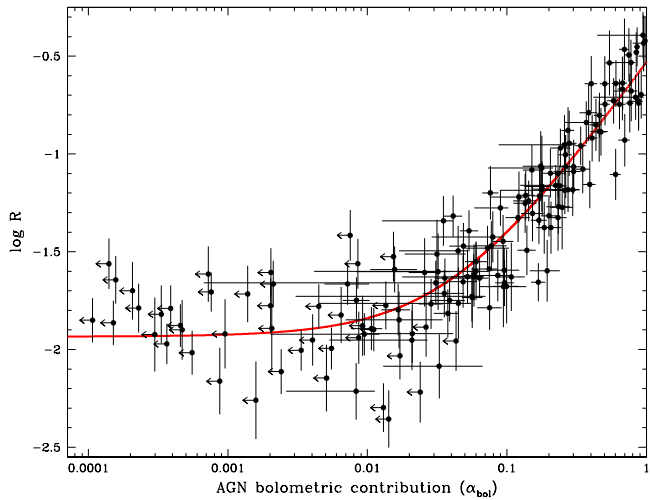}
\caption{Absorption-corrected ratio $R$ between the 6~$\mu$m and the bolometric luminosities, 
plotted against the AGN bolometric contribution $\alpha_\mathit{bol}$. The red solid curve is 
the best fit of equation~(\ref{eq}). We note that the absence of evident outliers suggests that 
our ULIRG sample encompasses a continuous series of SEDs, in which the variations can be entirely 
ascribed to the magnitude of the AGN contribution.} 
\label{np}
\end{figure}

\begin{figure}
\includegraphics[width=8.5cm]{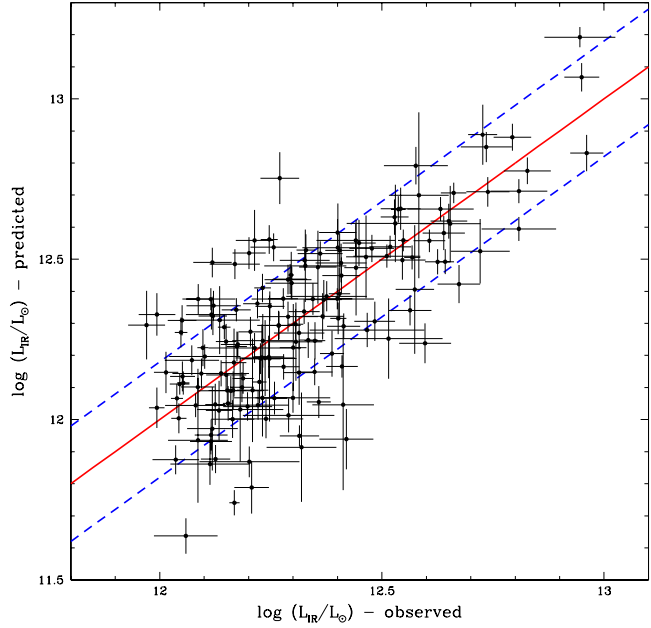}
\caption{Comparison between the total IR luminosity inferred from our model, assuming the best 
values of $R^\mathit{agn}$ and $R^\mathit{sb}$ as the true AGN/SB bolometric corrections, and 
the luminosity measured by \textit{IRAS} according to equation~(\ref{e1}). The blue dashed 
lines correspond to the 0.18~dex ($\sim$ 50 per cent) dispersion of the $R$--$\alpha_\mathit{bol}$ 
relation. Virtually all the entries fall within $\sim$0.3~dex from the exact match.} 
\label{pd}
\end{figure}

We now briefly review the basics of our diagnostic method, that has been presented and 
discussed in the previous papers of this series. At 5--8~$\mu$m the spectral properties 
associated to black hole accretion and star formation are widely different: the AGN exhibit 
a regular hot dust continuum, as opposed to the prominent pair of aromatic features typical 
of the SB galaxies. Moreover, only moderate dispersion is found within each class, making it 
possible to characterize the AGN and SB contribution to the emission of ULIRGs by means 
of spectral templates (Paper~II). Our model also takes into account 
the possible reddening of the AGN component due to a compact absorber along the line of sight; 
this localized extinction can not affect the SB component, which is much more diffuse and interspersed 
with the obscuring dust.\footnote{The effects of internal SB extinction are already included in the shape of the 
template, which is derived from observed SB spectra in the ULIRG luminosity range. Little difference 
is seen between our template and that obtained at lower luminosities by Brandl et al. (2006), which can be 
indeed interpreted as a result of the different internal obscuration (Paper~II; Rigopoulou et al. 1999).} Hence 
the degrees of freedom in our fitting procedure are the amplitudes of the AGN and SB templates and the optical 
depth to the AGN, which is supposed to follow the extinction law of Draine (1989). Apart from the flux normalization 
$f_6^\mathit{int}$, the basic parameters of our spectral decomposition are only the AGN contribution to the intrinsic 
(i.e. absorption-corrected) emission $\alpha_6$ and the optical depth $\tau_6$: 
\begin{equation}
f_\nu^\mathit{obs}(\lambda)=f_6^\mathit{int}\left[(1-\alpha_6)u_\nu^\mathit{sb}+
\alpha_6 u_\nu^\mathit{agn} e^{-\tau(\lambda)}\right], 
\end{equation}
where $u_\nu^\mathit{agn}$ and $u_\nu^\mathit{sb}$ are the AGN and SB templates. 
This model allows us to reproduce adequately the main features observed in the 
5--8~$\mu$m ULIRG spectra (see Fig.~\ref{apf}; all the spectra are available in the online publication, 
in order to illustrate their quality and the reliability of our spectral decomposition on the whole sample).
The results of the model fitting for each source are filed in Table~\ref{t1}. \\
We now introduce 
another useful diagnostic tool, namely the ratio between the 6~$\mu$m and the bolometric 
luminosities: since an AGN is much brighter around 6~$\mu$m than a SB of equal bolometric 
luminosity, this ratio is a straightforward indicator of the significance of nuclear 
activity within composite ULIRGs. Here we make use of the absorption-corrected ratio, defined 
as follows: 
\begin{equation}
R=\left(\frac{\nu_6 f_6^\mathit{int}}{F_\mathit{IR}}\right)
=\frac{R^\mathit{agn}R^\mathit{sb}}{\alpha_6 R^\mathit{sb}+(1-\alpha_6) R^\mathit{agn}},
\label{eq}
\end{equation}
where $R^\mathit{agn}$ and $R^\mathit{sb}$ are the intrinsic bolometric corrections for the 
separate AGN and SB components (we refer to Paper~I for the algebraic details). 
The $R$--$\alpha_6$ relation from equation~(\ref{eq}) has been superimposed to our absorption-corrected points, 
treating $R^\mathit{agn}$ and $R^\mathit{sb}$ as floating variables. The best fit yields 
$\log R^\mathit{agn}=-0.53 \pm 0.05$ and $\log R^\mathit{sb}=-1.93 \pm 0.02$, in excellent 
agreement with our previous estimates. Incidentally, this also proves that the possible 
effects of the $K$~correction are negligible, and that the average properties of the 
AGN and SB components do not seem to experience a strong evolution with redshift up to 
$z \sim 0.35$. It is now possible to obtain a quantitative estimate of the AGN contribution 
to the bolometric luminosity, by assuming $\kappa=R^\mathit{agn}/R^\mathit{sb}\sim$25: 
\begin{equation}
\alpha_\mathit{bol}=\frac{\alpha_6}{\alpha_6+\kappa(1-\alpha_6)}. 
\label{ab}
\end{equation}
The analytical steps described so far are summarized in Fig.~\ref{np}, in which the best 
fit of equation~(\ref{eq}) is shown as a function of $\alpha_\mathit{bol}$, that is  $R=\alpha_\mathit{bol}R^\mathit{agn}+(1-\alpha_\mathit{bol})R^\mathit{sb}$. The values 
of $\alpha_\mathit{bol}$ for the individual sources are listed in Table~\ref{t1}, with the 
1$\sigma$ confidence limits. \\
Concerning the 1~Jy ULIRGs, our results are in good agreement with those of Veilleux 
et al. (2009a): their ensemble and individual estimates are larger than ours by $\sim$10 per cent, 
but this seems to be a small systematic effect related to the AGN/SB \textit{zero points} (or, 
in other words, the factor $\kappa$). The work of Veilleux et al. (2009a) explores the connection 
between ULIRGs and quasars, and provides six different methods based on the \textit{Spitzer}-IRS spectra for 
computing the AGN contribution to the bolometric luminosity of both kinds of sources. The 
comparison among these six independent estimates gives a good measure of the uncertainties 
involved when considering the individual sources, which sometimes can be rather large with 
respect to the AGN contribution averaged over all methods. Such discrepancies can be 
regarded as the natural dispersion connected to the use of single indicators, which affects 
our narrow-band analysis as well. The scatter around the best fit of Fig.~\ref{np} is in fact 
significantly larger than the statistical uncertainty on the best values of $R^\mathit{agn}$ 
and $R^\mathit{sb}$. The actual 1$\sigma$ dispersion is 0.18~dex, nearly independent of 
$\alpha_\mathit{bol}$, and this should be considered the intrinsic dispersion of the 
6~$\mu$m to bolometric ratios for the AGN and SB components. An equivalent way 
of visualizing this point is shown in Fig.~\ref{pd}, where the total IR luminosities inferred from 
our spectral analysis, assuming the best values of $R^\mathit{agn}$ and $R^\mathit{sb}$ 
as the true bolometric corrections, are compared to the luminosities measured by \textit{IRAS} 
according to equation~(\ref{e1}). The natural dispersion is clearly brought out once again, 
and this limits the accuracy with which the AGN and SB components can be assessed in individual 
objects. \\
We finally remind what are the possible sources 
of systematic error in our approach: 1) the selection of a narrow wavelength range for 
our analysis prevents a complete understanding of the gas and dust properties, that could be 
better investigated by considering the whole \textit{Spitzer}-IRS spectra. 2) The use of AGN/SB 
templates is a strong assumption, and in some cases (less than 10 per cent) it turns out to be 
insufficient to reproduce the observed ULIRG spectra. This is due either to the presence of 
broad and irregular absorption features or to the flattening of the intrinsic AGN continuum. 
3) Different extinction curves can be involved instead of the power-law one that 
has been adopted here to model the AGN obscuration (e.g. Nishiyama et al. 2009, and 
references therein). This is related not only to physically different absorbers, but also to the 
accuracy of the screen approximation, and to the possible effects of radiative transfer, that 
have been neglected but can be important even at 5--8~$\mu$m. All these aspects have 
been taken into account and quantitatively constrained in Paper~II.

\section{Results and Discussion}

Our method proves to be very effective in detecting 
AGN components that are still invisible at other wavelengths (e.g. in the optical 
and/or hard X-ray domains). We obtain a detection rate of $\sim$70 per cent, with 113 detections 
out of 164 sources, which seems to represent the upper 
limit for the local ULIRG population. In the previous Section we have discussed the degree of 
uncertainty associated with the use of single (or narrow-band) diagnostics, which affects the 
extrapolation of the relative AGN/SB contribution to the bolometric luminosities. This 
notwithstanding, our  approach is very stable when applied to large samples, and 
allows a discussion on some general properties of ULIRGs. In this 
context we focus on the completeness and reliability of the optical classification, and 
review the correlation between nuclear activity and total IR luminosity, taking briefly 
into account also the possible connection with the morphological properties. Our results can 
be summarized as shown in Fig.~\ref{sr}, where the values of the AGN bolometric contribution 
$\alpha_\mathit{bol}$ are compared to the total luminosities and the optical classification. 

\subsection{Optical classification}

\begin{figure}
\includegraphics[width=8.5cm]{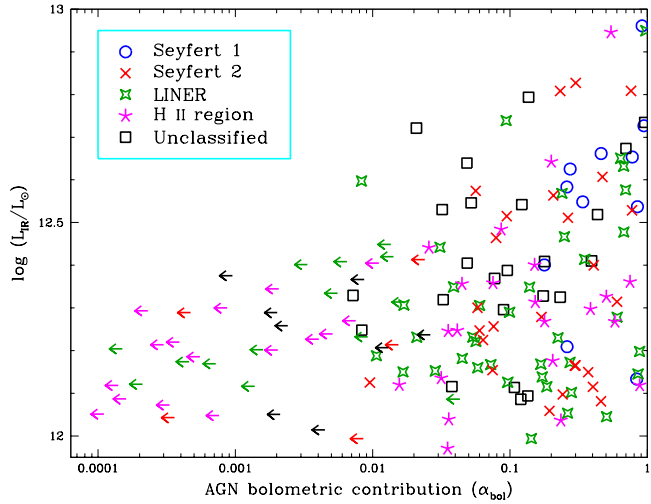}
\caption{The quantitative estimates of the AGN bolometric contribution $\alpha_\mathit{bol}$ 
(horizontal axis) as obtained from our analysis are plotted against the total IR luminosity 
of each source (vertical axis) and its spectral classification in the optical, according to 
the colour and symbol code defined in the box. The error bars have been omitted for clarity.}
\label{sr}
\end{figure}

\begin{figure}
\includegraphics[width=8.5cm]{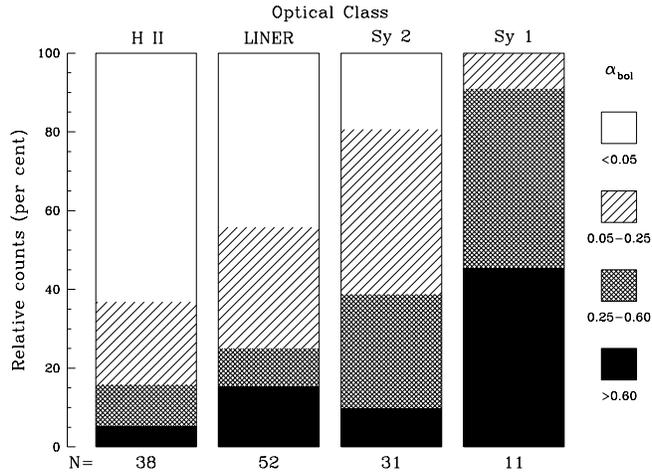}
\caption{Comparison between our values of $\alpha_\mathit{bol}$ and optical classification. 
Along the sequence of growing ionization from H~\textsc{ii} regions 
to type 1 Seyferts, as probed at optical wavelengths, the fraction of sources harbouring a significant 
AGN component (i.e. $\alpha_\mathit{bol}>0.25$) clearly evolves, and the incidence of SB-dominated 
objects drops as expected. The number of entries is shown below each class.}
\label{bo}
\end{figure}

\begin{figure}
\includegraphics[width=8.5cm]{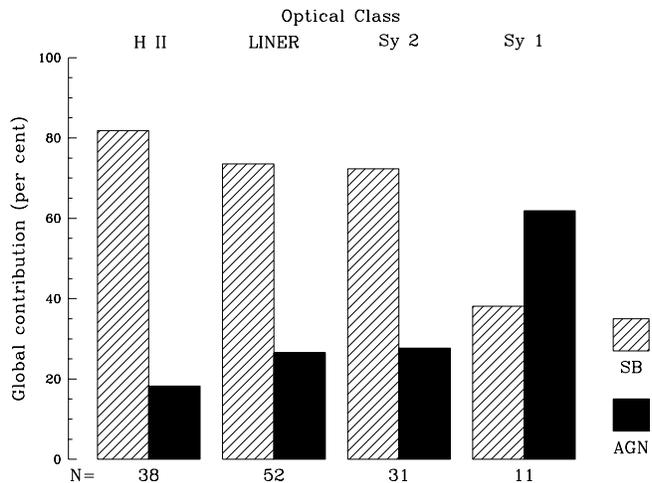}
\caption{The filled and shaded histograms represent respectively the global AGN and SB 
contributions integrated over all the sources in the different optical spectral classes. 
It is evident how optical spectroscopy has a great diagnostic value in the presence of 
broad emission lines, the key signatures of type 1 Seyfert galaxies. Conversely, whenever 
the degree of obscuration is higher, this kind of classification does not provide quantitative 
information. A good agreement still holds 
for the bare classification of Seyfert-like objects, but the average AGN content among 
type 2 Seyferts, LINERs and H~\textsc{ii} regions is actually rather similar.}
\label{glo}
\end{figure}

\begin{figure}
\includegraphics[width=8.5cm]{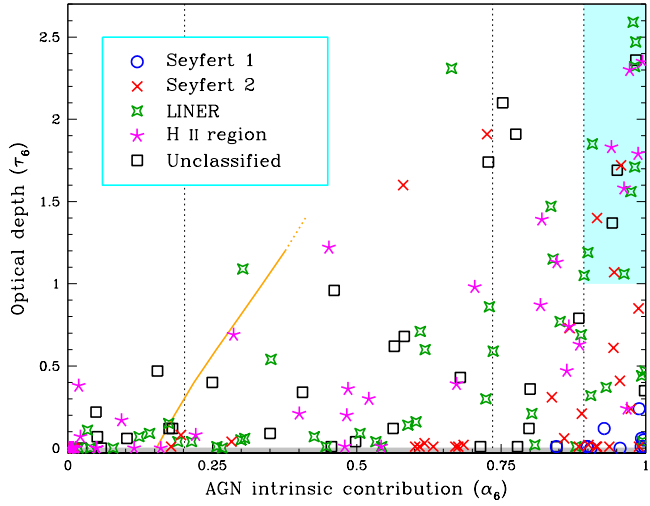}
\caption{Optical depth $\tau_6$ to the AGN component versus the absorption-corrected AGN contribution 
to the intrinsic 5--8~$\mu$m emission. Different colours indicate the optical classification, 
as shown in the box: the symbols are the same used in Fig.~\ref{sr}. The vertical lines refer to the 
bolometric scale, indicating respectively an AGN contribution to the overall IR emission of 1, 10 and 
25 per cent. The orange line represents our detection limit, due to the dispersion around the adopted 
AGN and SB templates. The exact position of this limit also depends on the signal-to-noise of the 
individual spectra, that in a few cases is rather low at 5--8~$\mu$m. For this reason the AGN detections 
falling next to this curve should be considered as tentative; we note that the bolometric 
contribution of such components is anyway negligible. The shaded region in the top right-hand 
corner encompasses the most interesting subclass of objects, those with significant but highly obscured 
nuclear activity ($\alpha_\mathit{bol}>0.25$ and $\tau_6 >1$). Since only three sources in this group 
show the typical signatures of Seyfert galaxies in their optical spectra, this area corresponds to the 
location of the \textit{elusive} AGN.}
\label{b2t}
\end{figure}

\begin{figure}
\includegraphics[width=8.5cm]{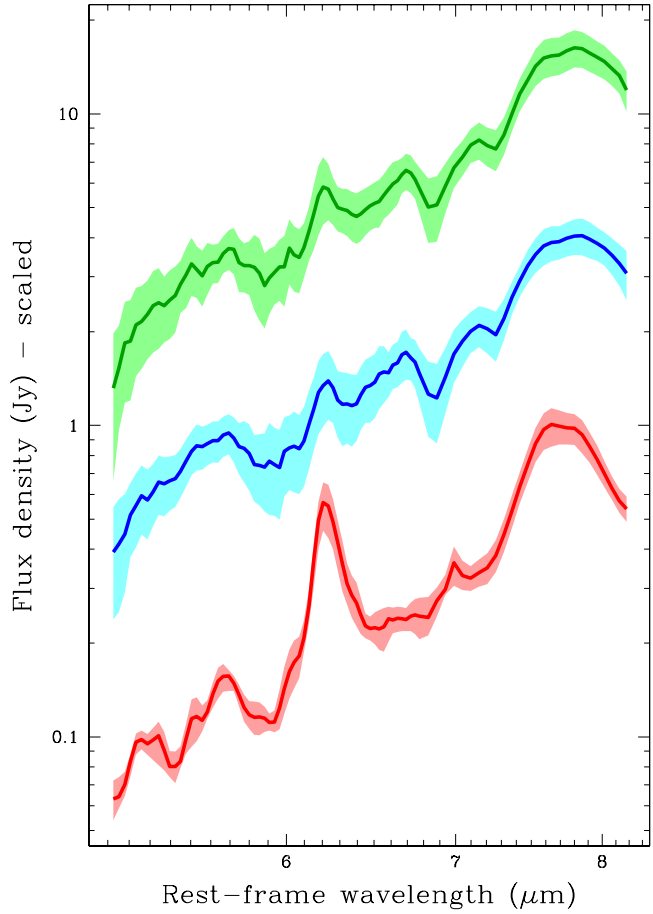}
\caption{Average 5--8~$\mu$m spectra of \textit{elusive} AGN ($\alpha_\mathit{bol}>0.25$, 
$\tau_6 >1$) compared with the typical SB emission. We have separated the sources 
that are optically classified as LINERs (in green) and those that are classified as H~\textsc{ii} 
regions (in blue): the spectral properties are anyway almost identical, and very different from 
the SB template (the average spectrum of five bright SB-dominated ULIRGs, in red). 
The shadings represent the 1$\sigma$ \textit{rms} dispersion.}
\label{el}
\end{figure}

Follow-up observations at visible wavelengths began soon afterwards the discovery of 
several extremely bright IR objects during the \textit{IRAS} mission (Houck et al. 1985). 
The availability of empirical diagnostics based on emission lines (e.g. Veilleux \& 
Osterbrock 1987) made it possible to explore the source of the ionizing radiation field 
in large and representative samples of IR galaxies at $L_\mathit{IR} > 10^{10} L_\odot$ 
(e.g. Armus, Heckman \& Miley 1989). Soon, it was clear that the systems with \textit{warm IRAS} 
colours ($f_{25}/f_{60}>0.2$; Sanders et al. 1988) display the typical properties of Seyfert 
galaxies and are the sites of dust-enshrouded nuclear activity. Also, objects of this class 
are found much more frequently as the total luminosity increases. \\
There are problems with optical diagnostics, though. The inner regions of ULIRGs 
are usually characterized by high obscurations at visible wavelengths, and extinction/reddening 
effects can reduce significantly the effectiveness of any emission-line criterion. In particular,
a number of objects end up with different classifications when considering 
different line ratios. This ambiguity affects mostly 
the objects classified either as H~\textsc{ii} regions or low-ionization nuclear emission-line 
regions (LINERs). The latter can actually be a different manifestation of the AGN family, possibly 
related to low accretion rates and/or low radiative efficiency (e.g. Maoz et al. 2005). On the 
other hand, the same (low) degree of ionization can be due to SB-driven thermal shocks and galactic 
winds (e.g. Sturm et al. 2006). The nature of LINERs is then controversial, and the ambiguity 
can not be resolved without more effective diagnostics (e.g. in the hard X-rays; 
Gonz{\'a}lez-Mart{\'i}n et al. 2009). This forbids a complete census of AGN activity among ULIRGs 
at optical wavelengths, even if some improvement can be obtained through a revision of 
the classification boundaries (e.g. Yuan, Kewley \& Sanders 2010). A further limitation of 
optical diagnostics is their poorly quantitative nature: it is difficult to correct the line ratios for 
extinction and to take into account the possible differential obscuration of the AGN and SB components. \\
A more detailed look into this issue is provided in Fig.~\ref{bo}, where the AGN bolometric contribution 
is plotted against the optical classification (i.e., here we collapse the plot of Fig.~\ref{sr} along the 
luminosity axis).
We have defined four different regions in the 
$\alpha_\mathit{bol}$ space with respect to the AGN \textit{weight}, that is: \textit{negligible} 
(region~1, $\alpha_\mathit{bol}<0.05$); \textit{minor} (region~2, $0.05<\alpha_\mathit{bol}<0.25$); 
\textit{significant} (region~3, $0.25<\alpha_\mathit{bol}<0.60$); \textit{dominant} 
(region~4, $\alpha_\mathit{bol}>0.60$). The separation values have been chosen as follows: 
$\alpha_\mathit{bol}=0.05$ represents 
a sort of 3$\sigma$ confidence limit, above which secure AGN detections are found. Hence all 
the SB-dominated sources and the more controversial cases fall into region~1 (beside 
many solid AGN detections). In our previous works $\alpha_\mathit{bol}=0.25$ was selected 
for the sake of a gross classification, corresponding to a luminosity ratio of 1:3 between the AGN 
and SB components (and, roughly, to the average AGN contribution). We keep this convention in 
the boundary between the region~2 and the region~3. Finally, we have chosen 
$\alpha_\mathit{bol}=0.60$ to single out the sources for which nuclear activity in the 
thermal IR is the dominant power supply: if we drop the assumption 
that $L_\mathit{bol} \simeq L_\mathit{IR}$ and consider the more realistic case 
$L_\mathit{bol} = 1.15$~$L_\mathit{IR}$ (as in Veilleux et al. 2009a), we find that 
region~4 is populated by the truly AGN-dominated ULIRGs ($L_\mathit{agn}>L_\mathit{sb}$). 
As expected, the fraction of SB-dominated objects shows a constant decline 
as the radiation field grows harder; in parallel, a clear upward evolution is seen in the 
frequency of sources harbouring a significant AGN component ($\alpha_\mathit{bol}>0.25$), 
i.e. the sources in which the contribution from nuclear activity to the energy budget becomes 
comparable to that from star formation. Concerning the individual classes, it is worth noting 
that a sizable number of very powerful AGN actually lies among LINERs (see also Risaliti, Imanishi \& Sani 2010). 
On the contrary, a handful of type 2 Seyfert-like sources are not seen at 5--8~$\mu$m: except for two cases 
(IRAS~09111$-$1007 and IRAS~18368+3549) these AGN components are actually detected, even if they 
are too close to the confidence limit. \\
Summarizing, a good agreement can be established between the general findings of optical and 
mid-IR diagnostics. In more quantitative terms, we have measured the global 
AGN/SB contribution to each of the optical spectral classes, integrating over all the corresponding 
entries; that is, we have computed $\alpha_\mathit{agn}=(\sum \alpha_i L_i)/(\sum L_i)$, with 
$\alpha_i$ and $L_i$ standing for the previous $\alpha_\mathit{bol}$ and $L_\mathit{IR}$. 
The results are shown in Fig.~\ref{glo}, a summary of which is also presented in Table~\ref{t2}. 
The optical classification of ULIRGs is conclusive only when the broad 
line region (BLR) can be probed, either directly or in polarized light (as usual for Seyfert 2 galaxies, 
e.g. Antonucci \& Miller 1985; Lumsden et al. 2001). In most cases, however, only narrow lines are 
available, and optical diagnostics of the individual sources can only provide incomplete 
(and sometimes misleading) information; in particular, reliable quantitative constraints to 
the AGN contribution can not be obtained when the BLR is not detected. The most striking 
indication of Fig.~\ref{glo} is that the AGN content among type 2 Seyferts, 
LINERs and H~\textsc{ii} regions is very similar, as well as the median mid-IR spectral properties 
(however with a different scatter, see below). 
The key role of dust obscuration is shown in Fig.~\ref{b2t}, where the amount of reddening 
$\tau_6$ is plotted against the AGN contribution to the intrinsic 5--8~$\mu$m emission and 
the optical type. Seyfert 1 objects are found in the bottom right-hand corner, 
indicating negligible obscuration and dominant AGN contribution. All the other classes 
are characterized by a large and somewhat unexpected scatter, that in the case of H~\textsc{ii} 
regions is actually dramatic. The shaded region in the top right-hand corner of the plot 
encompasses the most intriguing subclass of objects, those harbouring significant but 
highly obscured nuclear activity ($\alpha_\mathit{bol}>0.25$ and $\tau_6 >1$). Among the 
20 entries, only three type 2 Seyferts are found, in contrast with the nine LINERs and five 
H~\textsc{ii} regions. This area therefore corresponds to the location of the elusive AGN. 
This is confirmed in Fig.~\ref{el}, where the 5--8~$\mu$m spectral properties of these 
objects are shown against the expected emission originating from pure star formation 
activity. Aromatic features are strongly suppressed, diluted into the intense hot dust 
continuum. The signatures of dense absorbers along the line of sight are evident, implying 
the presence of a deeply obscured and compact energy source. Moreover, this rules out the 
possibility that the PAH suppression is due to the young age of the stellar burst (e.g. 
Efstathiou, Rowan-Robinson \& Siebenmorgen 2000), which is not even supported by the 
morphological properties. 

\subsection{Trend with luminosity}

\begin{figure}
\includegraphics[width=8.5cm]{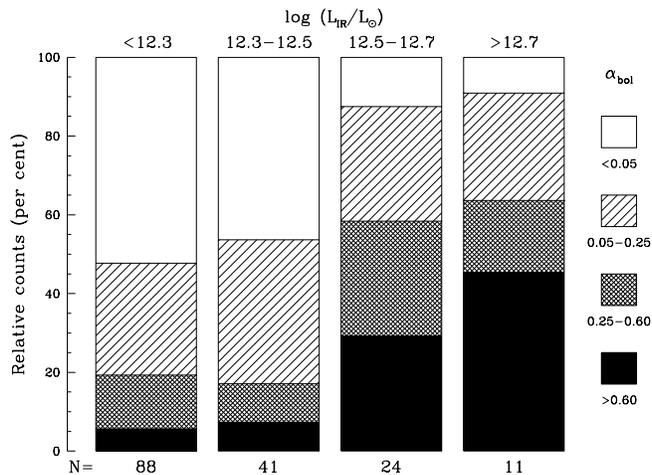}
\caption{Distribution of the AGN bolometric contributions within the different intervals 
of the ULIRG luminosity range. The difference between the high luminosity and the low luminosity 
bins hints at a sudden change in the nature of the energy source around 
$L_\mathit{IR} \simeq 3 \times 10^{12} L_\odot$.}
\label{bir}
\end{figure}

\begin{figure}
\includegraphics[width=8.5cm]{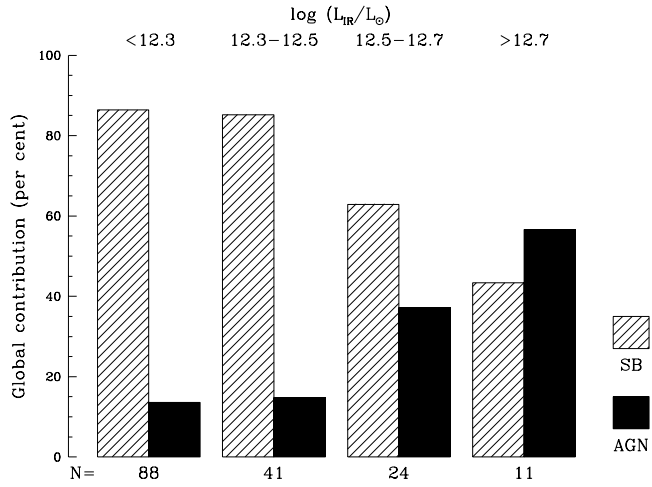}
\caption{As in Fig.~\ref{glo}. Here the relative AGN/SB contribution is summed 
over all the sources in the different bins of IR luminosity. A sharp increase of the 
AGN fraction is observed around $\log (L_\mathit{IR}/L_\odot) \sim 12.5$, and the 
trend of a nuclear activity increasing with the total luminosity is clearly established.}
\label{tr}
\end{figure}

The availability of a reliable measure of the relative AGN/SB contribution to ULIRGs allows 
a quantitative investigation of the relation between total luminosity and AGN contribution. 
As mentioned above, it is known from optical spectroscopy that the fraction of Seyfert-like 
systems among IR galaxies grows along with luminosity (Veilleux et al. 1995; Kim, 
Veilleux \& Sanders 1998; Veilleux et al. 1999; Goto 2005). 
The existence of a physical trend has been confirmed at mid-IR wavelengths by a wealth of 
studies, before and after the advent of \textit{Spitzer} (e.g. Tran et al. 2001; Imanishi 2009). 
In order to check how our findings fit into this scheme, we first divide the ULIRG luminosity 
range in four different intervals, whose width is either 0.2 or 0.3~dex, not strictly constant but 
larger (or comparable, at worst) to the expected dispersion of our measures. This assumption 
does not bias the discussion and allows a simple comparison with previous works. 
One can easily appreciate from Fig~\ref{bir} how star formation and nuclear activity are the primary 
engine at the opposite ends of the ULIRG luminosity range. The SB component dominates 
at $\log (L_\mathit{IR}/L_\odot)<12.5$, where the AGN is a significant contributor 
in only 1/5 of the cases, i.e. in $\sim$30 per cent of the 
composite sources. The power supplied by black hole accretion grows stronger along 
the luminosity scale, and ultimately it represents the trigger of the extreme IR 
activity. The AGN/SB energy balance is shown in Fig.~\ref{tr} and summarized in 
Table~\ref{t2}. In conclusion, the selection of a suitable wavelength range for 
ULIRG diagnostics allows us to quantify the AGN/SB contribution in the largest sample 
of local sources collected so far, confirming and enlarging all the previous results on the 
correlation between nuclear activity and IR luminosity (see also Imanishi et al. 2008; Veilleux 
et al. 2009a; Imanishi, Maiolino \& Nakagawa 2010). \\
Recently, Valiante et al. (2009) have adopted a similar decomposition method to derive the 
AGN/SB contribution to IR galaxies of lower luminosity, and proposed a backward evolution 
model to interpret the number counts at high redshift. Their sample includes also $\sim$35 
ULIRGs. By using a narrow window around the 6.2~$\mu$m aromatic feature, these authors measure 
the ratio between the 6~$\mu$m luminosity of the AGN component and the total IR luminosity.  
Interestingly, only a trend in the AGN detection rate is found among LIRGs (the luminous IR 
galaxies, $L_\mathit{IR} > 10^{11} L_\odot$) as a hint to the change in the intrinsic distribution of $\alpha_\mathit{bol}$. 
The AGN trend can be anyway recovered also at $L_\mathit{IR} < 10^{12} L_\odot$ through 
Monte Carlo simulations, intended to reproduce the properties of the data (detection rate, mean 
and standard deviation) in each bin of luminosity. Translated into our formalism, it follows from Valiante et al. (2009) that 
$R \alpha_6 e^{-\tau_6} \propto (L_\mathit{IR})^\gamma$, where $\gamma = 1.4 \pm 0.6$. We find 
a consistent but slightly softer dependence, with $\gamma_\mathit{best} \approx 0.7-1.1$ (not substantially 
modified when applying the absorption correction, i.e. considering $R \alpha_6$ only, due to the large scatter). 
If the discrepancy is real, we argue that a single power law may not adequately fit the distribution over $\sim$2 
orders of magnitude in luminosity. On average, LIRGs can not be considered a true scaled version of 
ULIRGs, as suggested e.g. by the different fraction of interacting systems (see Yuan et al. 2010 for the 
most recent and comprehensive analysis on this subject), when interactions are fundamental in order to start 
the ultraluminous phase. Hence, there can be \textit{hidden} discontinuities along the IR luminosity range. 
From both Fig.~\ref{bir} and Fig.~\ref{tr}, for instance, the AGN trend is suggested not to be smooth; instead, 
a sharp turning point from star formation to nuclear activity seems to occur around 
$L_\mathit{IR} \simeq 3 \times 10^{12} L_\odot$. Above this threshold, extreme IR luminosity 
can not be explained without a strong AGN contribution. In different terms, this suggests 
the possibility of a maximum SB luminosity. A limiting luminosity can indeed be associated to 
SB activity through an Eddington-like argument as $L_{limit}=(4 f_g c/G) \sigma^4$, where 
$f_g$ is the gas mass fraction and $\sigma$ the stellar velocity dispersion (Murray, Quataert 
\& Thompson 2005). The momentum injection in the interstellar medium due to the star formation 
events (and nuclear accretion) is able to drive a significant fraction of the gas out of the 
central regions, giving rise to a self-regulation mechanism that prevents further activity. 
The gas displacement is thought not to affect the smaller scales, and can even enhance the fuelling 
of the black hole until the latter reaches its own Eddington limit. It is worth noting that assuming 
$f_g=1/6$ (Downes \& Solomon 1998) and $\sigma=161$~km~s$^{-1}$ (Dasyra et al. 2006b) we obtain 
$L_{limit} \sim 5 \times 10^{12} L_\odot$, a value that is positioned just above the typical SB 
luminosity in local ULIRGs. However, 
the interplay and mutual feedback between the AGN and SB components during a merger process 
are still debated (e.g. Springel et al. 2005; Johansson, Naab, \& Burkert 2009). 

\subsection{Morphological properties}

All the optical and near-IR imaging surveys reveal a tight connection between extreme IR 
luminosity and large-scale gravitational disturbance: above $L_\mathit{IR}>10^{12} L_\odot$ 
virtually all the systems appear to be involved in a different stage of an interaction or 
merger process (Kim et al. 2002, and references therein). As briefly mentioned above, 
numerical studies have been very useful to shed light on the dynamics of these major 
encounters: the huge gas and dust content observed in the central regions of ULIRGs has 
been funneled via angular momentum dissipation, and act as both a reservoir and an absorbing screen 
for black hole accretion and star formation. There is an obvious link to the AGN/SB feedback. 
It was early suggested that \textit{warm} ULIRGs could represent the transition stage between 
cooler systems and optically bright quasars (Sanders et al. 1988). According to this scenario, 
the radiation pressure, the violent stellar winds and the supernova-driven shocks pervading 
the nuclear environment eventually expel the gas and the dust from the line of sight to the 
active nucleus, as the SB starts to fade. When the bulk of the dust layers responsible for 
the reprocessing are swept away, optical and UV photons are able to escape, unveiling the 
quasar (Hopkins et al. 2006, and references therein). Some recent works, however, 
seem to exclude the possibility that the ULIRG population as a whole evolve into powerful optical 
quasars. The properties of the massive merger remnants look similar to those of quiescent elliptical 
galaxies, and the masses inferred for the ULIRG black holes lie in the Seyfert range. The luminosity 
of ULIRGs is suggested to be quasar-like only because of the fuelling mechanism, that allows 
both star formation and accretion to radiate at near-Eddington efficiencies (Tacconi et al. 2002; Murray 
et al. 2005; Dasyra et al. 2006b). This latter explanation is still in part controversial, due to the 
large discrepancies among the different methods adopted to determine the black hole mass. There is 
opposite evidence of a tight overlap between radio-quiet PG quasars and highly-evolved ULIRGs in 
terms of both black hole mass and accretion rate (Veilleux et al. 2006; Veilleux et al. 2009b). 
However, such a similarity could be due also to the lack of a further evolutionary stage linking 
ULIRGs to optical quasars. \\
The study of these problems goes beyond the scope of the present work, and requires 
additional diagnostics, e.g. the optical depth of the 9.7~$\mu$m 
silicate feature. This can be plotted against the strength of PAH emission in order to 
probe both the age of the SB and the geometrical structure of the dust. The resulting 
diagram gives not only a direct classification, but also possible hints at the 
evolutionary path of ULIRGs (Spoon et al. 2007). It is anyway worth noting that 
also our analysis confirms the presence of a larger nuclear activity among the 
advanced mergers. Morphological information is available for almost 90 per cent of the 
sources in our sample. In the present context, we have adopted a conventional distance 
to separate early and advanced merger stages. In the latter class we include the single nuclei, 
the remnants and the binary (or multiple) systems up to a projected distance of $\sim$5~kpc; 
the former class comprises the wider binary systems, with distinct but clearly interacting 
nuclei (see Table~\ref{t1}). The global AGN contribution to the 94 compact systems 
 turns out to be $\sim$32 per cent, while it reaches up to only $\sim$20 per cent in 
the 47 loose systems (and also in the 23 objects with no imaging information). In spite of the 
uncertainties in our classification,\footnote{The apparent nuclear separation does not 
only depend on the merger stage but also on the inclination of the merger plane, which is implicitly 
assumed to be random. We also stress that in general the physical distance is not a monotonic 
function of time.} this simple test supports the possibility of an evolution in the nature of the energy 
source with the proceeding of the interaction (see also Farrah et al. 2009b).

\section{Evolution with redshift}

\begin{figure}
\includegraphics[width=8.5cm]{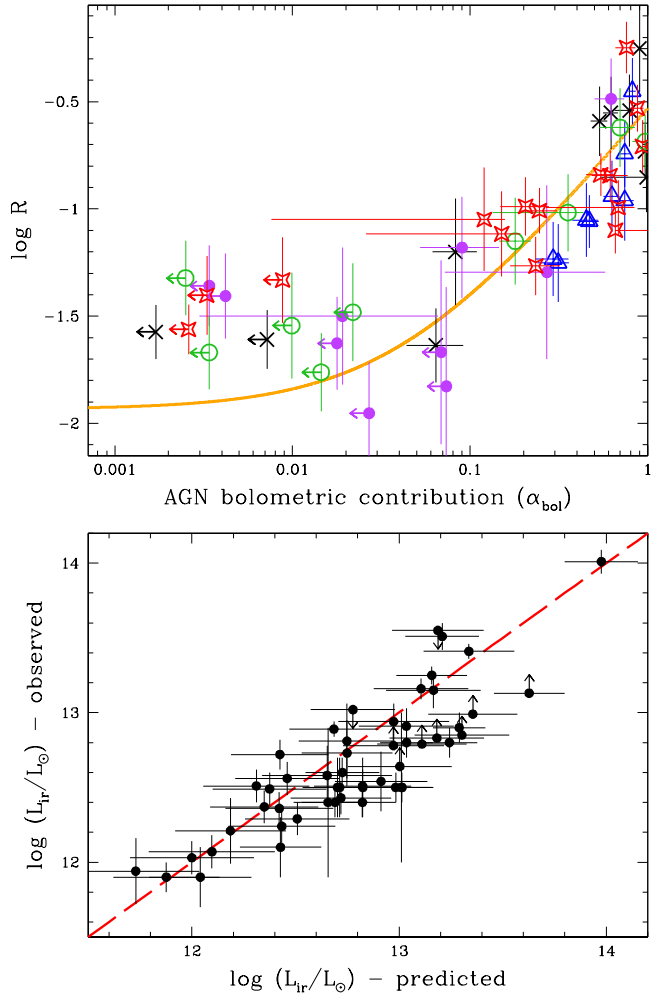}
\caption{(a) Distribution in the $R$--$\alpha_\mathit{bol}$ diagram of ULIRG-like systems 
at $z \sim 1$ with respect to the best fit of the local sample (orange curve). Different 
colours and symbols are used for the different selections: red diamonds (Hern{\'a}n-Caballero 
et al.~2009), violet points (Farrah et al.~2009a), green rings (Brand et al.~2008), 
blue triangles (Stanford et al.~2000), black crosses (other selection; see also Table~\ref{t3}). 
The agreement suggests only minor changes with redshift of the SED large-scale shape  
for the AGN and SB components. (b) This is also confirmed by the good match between the IR 
luminosities predicted in a non-evolutionary scenario (see the text for details) and those 
available from broad band photometry in the literature.}
\label{hz}
\end{figure}

The extension of our diagnostic method at higher redshift requires some 
caution. The use of AGN/SB templates is based on sound observational evidence, 
but its validity has to be tested in different regions of the redshift/luminosity space. 
In principle an evolution in both directions is expected, and moderate variations of 
the AGN/SB spectral shapes are actually observed in the local Universe as a 
function of luminosity. IR galaxies at $z \sim 1-3$ appear to be powered by the 
same physical process characterizing local ULIRGs, that is a merger-driven 
combination of intense star formation and nuclear accretion (Dasyra et al. 2008; 
Melbourne et al. 2009). The question is whether the similarity holds also in terms 
of spectral properties. The $R$--$\alpha_\mathit{bol}$ test can be 
useful to investigate this point. Again, it is important to define a solid measure 
of $L_\mathit{IR}$ for the fainter objects. Concerning local sources, the \textit{IRAS} 
flux density $f_{60}$ is known to be a good indicator of the total IR luminosity, so 
that $\nu L_\nu$~(60~$\mu$m) is sometimes adopted to obtain an alternative definition 
of the ULIRG class. By estimating $L_\mathit{IR}$ in this way in our sample we indeed 
find an excellent agreement (within 0.1~dex) with equation~(\ref{e1}). 
Hence the rest-frame luminosity at $\sim$50~$\mu$m is a reliable proxy of $L_\mathit{IR}$ 
at $z < 0.35$. We can argue how far backward this fiducial point can be shifted in wavelength. 
In Paper~II we have discussed the enhancement of the AGN over SB brightness 
ratio at 5--8~$\mu$m due to the hot dust component; we have also shown how rapidly it 
declines and completely vanishes at $\sim$30~$\mu$m. In this view, the 70~$\mu$m fluxes 
provided by the Multiband Imaging Photometer onboard \textit{Spitzer} (MIPS; Rieke et 
al. 2004) allow to assess the luminosity of IR galaxies at $z \sim 1$ regardless of their power 
source. Our basic assumption is that the local toolbox (AGN/SB templates and bolometric 
corrections, extinction law, 30~$\mu$m rest-frame to IR ratio) can be used as a fair approximation in the 
analysis of the distant populations. \\
We have first applied a filter with narrow gaussian profile to 120 sources in 
our main sample at $z < 0.2$, in order to derive their monochromatic 30~$\mu$m rest-frame 
luminosity and check the correlation with the total energy output, which turns out to be very good. 
A near proportionality holds, and the best fit is obtained for 
$F_\mathit{IR} \propto (f_{30\mu m})^{0.93}$. We have then searched in the literature 
for faint IR galaxies whose MIPS $f_{70}$ (or \textit{IRAS} $f_{60}$) flux allows a 
good sampling of the 30~$\mu$m rest-frame band, and found 52 ULIRG-like systems at 
$0.5 < z < 1.5$. The 30~$\mu$m rest-frame luminosities have been extrapolated from the measured 
fluxes by assuming a power-law trend for the continuum, whose spectral index 
$\Gamma_{30\mu m} = 3.37 \pm 0.82$ is the average over the local sources. The selection 
of the 52 objects is very heterogeneous, and the lack of completeness precludes the study 
of the AGN detection rate and global contribution. On the other hand, the great variety 
is a desirable feature for a control sample. A few known hyperluminous IR galaxies 
(HLIRGs, $L_\mathit{IR}>10^{13} L_\odot$) and quasar-like objects 
are also included: by applying our deconvolution model we are able to obtain an adequate 
fit in every case. The properties of the sources, their parent samples and the results of our 
analysis are summarized in Table~\ref{t3}. \\
The location in the $R$--$\alpha_\mathit{bol}$ diagram of the high-redshift IR galaxies 
is shown in Fig.~\ref{hz}(a). Their distribution is in good agreement with the best fit 
for \textit{local} ULIRGs. We stress that this result does not come from a circular argument. 
The $R$--$\alpha_\mathit{bol}$ test employs multiple 
diagnostics: the 5--8~$\mu$m AGN/SB spectral shapes, the AGN/SB ratios between the 
5--8~$\mu$m and the 8--1000~$\mu$m emission, the correlation between the 30~$\mu$m 
and the total ULIRG luminosity. If any of these elements were a strong 
function of redshift, evidence for either an ensemble deviation or 
dramatic outliers would be found. It is also worth noting, as in Fig.~\ref{hz}(b), 
that our estimates of $L_\mathit{IR}$ (reconstructed from a single far-IR point) are well 
matched to the tabulated values, that are computed from a broader band photometry, 
even if still limited. In more detail, the distribution of the high-redshift 
entries in the $R$--$\alpha_\mathit{bol}$ plot suggests a small change of $R^\mathit{sb}$, 
the bolometric correction for SB-dominated sources. There are two possible explanations 
for this effect: 1) a missed AGN detection, due either to the bad quality of the single 
spectra or to a modification of the AGN/SB templates; 2) an underestimate of 
$L_\mathit{IR}$, since the 30~$\mu$m flux does not properly represent the typical 
dust temperature of a SB environment. The latter argument is perhaps the most 
likely. \\
Evidence against dramatic spectral variations has been found also around $z \simeq 2.3$, by 
applying our AGN/SB decomposition to the stacked spectra of 24~$\mu$m-selected sources 
and submillimetre galaxies (Watabe et al. 2009). The assumption of AGN/SB templates and 
bolometric corrections similar to the local ones leads to an average AGN content which 
is fully consistent with the main properties of both populations (Sajina et al. 2007; 
Men{\'e}ndez-Delmestre et al. 2009). We conclude that, as a first approximation, 
the SED large-scale properties of ULIRG-like systems are not subject to significant 
evolution with redshift. This hints to a variant of our diagnostics, where the fitting of templates to 
the mid-IR spectra is replaced with the measurements of mid-IR colours or spectral slopes. 
With the advent of \textit{JWST} and \textit{Herschel}, the combined spectral and photometric 
coverage will enable the measure of both the 3--8~$\mu$m rest-frame slope (see also Risaliti et 
al. 2010) and the bolometric correction: a simple $\Gamma$--$R$ diagnostic diagram will then 
provide the quantitative analysis of much fainter IR sources in the deep fields.

\section{Conclusions}

The \textit{Spitzer}-IRS unprecedented sensitivity allowed a deeper investigation of the 
role of supermassive black hole accretion and intense star formation as the engine underlying 
extreme IR activity. In particular, the 5--8~$\mu$m rest-frame wavelength range has proven to be very 
suitable for solid diagnostics of ULIRGs: the large difference in this band between the typical spectral 
signatures of SB galaxies and AGN makes it possible to fully characterize both components 
and disentangle their contribution. In this paper we have presented the application of our spectral 
decomposition method to the largest sample of local ULIRGs studied so far, consisting of 164 
sources up to $z \simeq 0.35$. The size of our sample, which is not affected by any significant 
bias with respect to the nature of the energy source, allowed us to obtain the following results: \\
1) Our method has proven to be very effective in the discovery of faint/obscured AGN components, 
with an AGN detection rate among ULIRGs of $\sim$70 per cent (113/164 secure detections, 
plus a dozen of ambiguous cases among the fainter objects). This rate is comparable to that achieved 
by collecting together all the other multiwavelength diagnostics. \\
2) In terms of global contribution, star formation is confirmed as the dominant power source for extreme 
IR activity: only $\sim$27 per cent of the total luminosity of our sample turns out to have a gravitational 
origin. Nevertheless, even if they are usually minor contributors with respect to the concurrent SB events, 
AGN play a key role in the ULIRG phenomenon, since obscured nuclear activity is actually in place in 
many sources that are optically classified as H~\textsc{ii} regions (or LINERs) as well. Elusive AGN 
components are important in $\sim$10 per cent of ULIRGs. \\
3) The coverage of the entire ULIRG luminosity range allows a comprehensive and \textit{quantitative} 
re-analysis of the well-known correlation 
between the relative contribution of nuclear activity and the overall energy output. The increasing 
trend as a function of the total IR luminosity is clearly recovered: the average AGN contribution is almost 
negligible at the lower luminosities, but it gradually grows and eventually outshines the SB counterpart.
A physical turning point is suggested to exist around $L_\mathit{IR} \simeq 3 \times 10^{12} L_\odot$, 
possibly related to the complex interplay between the two processes at work. \\
4) As for the morphological properties, the AGN content turns out to be larger in 
correspondence of a more advanced merger stage; this supports the possibility of a time dependence in the relative 
AGN/SB contribution to the luminosity of interacting systems, consistent with the evolutive scenario 
according to which some ULIRGs may represent an intermediate phase in the formation of optically bright 
quasars. \\ 
5) The analysis of a control sample of 52 IR-luminous sources at $z \sim 1$ shows that to 
first approximation the spectral properties and the large-scale SED shapes of the AGN and SB 
components do not suffer substantial evolution with redshift. Our method then provides also an initial 
measure of the role of black hole accretion and star formation among ULIRG-like systems at earlier 
cosmic epochs. Moreover, the large spectral coverage that will be achieved with the upcoming IR 
facilities hints at the photometric extension of our diagnostic technique, based on accessible indicators 
such as the $\sim$5~$\mu$m rest-frame slope and the bolometric correction.

\section*{Acknowledgments}

We are grateful to the anonymous referee for the constructive comments and suggestions. 
This work has made use of the NASA/IPAC Extragalactic Database (NED) which is operated 
by the Jet Propulsion Laboratory, California Institute of Technology, under contract with 
the National Aeronautics and Space Administration. We acknowledge financial support from 
NASA grant NNX09AT10G and ASI-INAF I/088/06/0 contract.



\clearpage
\thispagestyle{empty}
\begin{landscape}
\begin{table*}
\caption{General properties and spectral parameters of our local ULIRG sample. 
(1)~\textit{IRAS} name, (2)~redshift, (3)~IR luminosity, (4)~AGN bolometric 
contribution~(in per cent), (5)~1$\sigma$ confidence range for $\alpha_\mathit{bol}$, 
(6)~AGN optical depth, (7)~\textit{Spitzer} programme, (8)~optical class, 
(9)~morphology (Compact/Loose), (10)~references for optical and morphological type.}
\label{t1}
\begin{scriptsize}
\begin{tabular}{ccccccccccccccccccccc}
\hline \hline
\\
Object & $z$ & $\log (L_\mathit{IR}/L_\odot)$ & $\alpha_\mathit{bol}$ & $\Delta \alpha$ & $\tau_6$ & PID 
& Type & IC & Refs. & &
Object & $z$ & $\log (L_\mathit{IR}/L_\odot)$ & $\alpha_\mathit{bol}$ & $\Delta \alpha$ & $\tau_6$ & PID 
& Type & IC & Refs. \\
(1) & (2) & (3) & (4) & (5) & (6) & (7) & (8) & (9) & (10) & & 
(1) & (2) & (3) & (4) & (5) & (6) & (7) & (8) & (9) & (10) \\
\\
\hline
00091$-$0738 & 0.118 & 12.27$\pm$0.03 & 58 & 52$-$64 & 2.30 & 3187 & H~\textsc{ii} & C & 1,15 & & 
09039+0503 & 0.125 & 12.16$\pm$0.07 & 5.8 & 4.9$-$6.9 & 0.71 & 3187 & LINER & C & 1,17 \\
00182$-$7112 & 0.327 & 12.95$\pm$0.04 & 98 & 96$-$99 & 0.47 & 666 & LINER & C & 2,16 & & 
09111$-$1007 & 0.054 & 12.04$\pm$0.02 & $-$ & $<0.05$ & $-$ & 30323 & Sy 2 & C & 3 \\
00188$-$0856 & 0.128 & 12.41$\pm$0.04 & 35 & 31$-$39 & 0.37 & 105 & LINER & C & 1,17 & & 
09116+0334 & 0.145 & 12.17$\pm$0.06 & $-$ & $<0.09$ & $-$ & 2306 & LINER & L & 1,15 \\
00199$-$7426 & 0.096 & 12.33$\pm$0.02 & 0.7 & $<1.1$ & 0.47 & 105 & $-$ & L & 16 & & 
09320+6134 & 0.039 & 11.99$\pm$0.02 & 14 & 11$-$16 & $<0.02$ & 105 & LINER & C & 10,22 \\
00275$-$0044 & 0.242 & 12.53$\pm$0.11 & 3.2 & 0.8$-$3.9 & $<0.01$ & 105 & $-$ & $-$ & $-$ & & 
09425+1751 & 0.128 & 12.08$\pm$0.07 & 46 & 40$-$53 & 0.41 & 30121 & Sy 2 & C & 10,21 \\
00275$-$2859 & 0.278 & 12.65$\pm$0.07 & 78 & 59$-$86 & $<0.24$ & 105 & Sy 1 & L & 4,18 & & 
09463+8141 & 0.156 & 12.31$\pm$0.04 & 1.5 & $<2.0$ & $<0.06$ & 105 & LINER & C & 1,15 \\
00397$-$1312 & 0.262 & 12.95$\pm$0.08 & 54 & 50$-$58 & 0.24 & 105 &  H~\textsc{ii} & C & 1,17 & & 
09539+0857 & 0.129 & 12.10$\pm$0.08 & 28 & 25$-$32 & 1.85 & 3187 & LINER & C & 1,17 \\
00406$-$3127 & 0.342 & 12.81$\pm$0.06 & 76 & 71$-$81 & 0.85 & 105 & Sy 2 & C & 5,16 & & 
10036+2740 & 0.165 & 12.32$\pm$0.08 & 3.3 & $<6.7$ & $<0.96$ & 30407 & $-$ & C & 15 \\
00456$-$2904 & 0.110 & 12.22$\pm$0.03 & $-$ & $<0.05$ & $-$ & 3187 & H~\textsc{ii} & L & 1,17 & & 
10091+4704 & 0.246 & 12.65$\pm$0.04 & 64 & 57$-$71 & 2.59 & 105 & LINER & C & 12,15 \\
00482$-$2720 & 0.129 & 12.09$\pm$0.07 & 2.0 & $<2.6$ & $<0.04$ & 3187 & LINER & L & 1,15 & & 
10190+1322 & 0.077 & 12.05$\pm$0.03 & $-$ & $<0.02$ & $-$ & 3187 & H~\textsc{ii} & L & 1,15 \\
01003$-$2238 & 0.118 & 12.33$\pm$0.06 & 50 & 46$-$54 & 1.58 & 105 & H~\textsc{ii} & C & 1,17 & & 
10339+1548 & 0.197 & 12.41$\pm$0.07 & 1.8 & $<5.9$ & $<1.60$ & 30407 & Sy 2 & $-$ & 7 \\
01166$-$0844 & 0.188 & 12.12$\pm$0.06 & 88 & 78$-$94 & 2.35 & 3187 & H~\textsc{ii} & L & 1,15 & & 
10378+1109 & 0.136 & 12.35$\pm$0.04 & 14 & 12$-$16 & 0.21 & 105 & LINER & C & 1,15 \\
01199$-$2307 & 0.156 & 12.31$\pm$0.06 & 15 & 12$-$18 & 1.39 & 105 & H~\textsc{ii} & L & 1,15 & & 
10485$-$1447 & 0.133 & 12.22$\pm$0.07 & 5.7 & 4.7$-$6.7 & 0.16 & 3187 & LINER & L & 1,15 \\
01298$-$0744 & 0.136 & 12.36$\pm$0.05 & 75 & 71$-$80 & 1.79 & 105 & H~\textsc{ii} & C & 1,15 & & 
10494+4424 & 0.092 & 12.20$\pm$0.03 & $-$ & $<0.02$ & $-$ & 2306 & LINER & C & 12,15 \\
01355$-$1814 & 0.192 & 12.48$\pm$0.05 & 8.6 & 6.7$-$11 & 0.98 & 105 & H~\textsc{ii} & L & 1,15 & & 
10558+3845 & 0.208 & 12.24$\pm$0.07 & 1.7 & $<3.0$ & $<0.34$ & 20589 & $-$ & C & 20 \\
01388$-$4618 & 0.090 & 12.12$\pm$0.03 & 1.6 & $<2.3$ & 0.69 & 666 & H~\textsc{ii} & C & 6,16 & & 
10565+2448 & 0.043 & 12.05$\pm$0.01 & $-$ & $<0.09$ & $-$ & 105 & H~\textsc{ii} & L & 10,22 \\
01494$-$1845 & 0.158 & 12.29$\pm$0.05 & $-$ & $<0.3$ & $-$ & 105 & $-$ & C & 15 & & 
11029+3130 & 0.199 & 12.42$\pm$0.06 & $-$ & $<1.7$ & $-$ & 30407 & LINER & C & 12,15 \\
01562+2527 & 0.166 & 12.24$\pm$0.08 & $-$ & $<0.6$ & $-$ & 30407 & H~\textsc{ii} & $-$ & 7 & & 
11095$-$0238 & 0.107 & 12.28$\pm$0.03 & 60 & 56$-$64 & 1.56 & 105 & LINER & C & 1,17 \\
01569$-$2939 & 0.141 & 12.27$\pm$0.05 & 18 & 15$-$21 & 1.13 & 2306 & H~\textsc{ii} & C & 1,15 & & 
11119+3257 & 0.189 & 12.66$\pm$0.03 & 46 & 42$-$50 & $<0.01$ & 105 & Sy 1 & C & 12,15 \\
01572+0009 & 0.163 & 12.63$\pm$0.03 & 27 & 17$-$33 & $<0.01$ & 105 & Sy 1 & C & 1,17 & & 
11130$-$2700 & 0.136 & 12.14$\pm$0.06 & 17 & 15$-$20 & 1.15 & 2306 & LINER & C & 1,15 \\
02021$-$2104 & 0.116 & 12.09$\pm$0.04 & 12 & 10$-$15 & 1.91 & 3187 & $-$ & C & 17 & & 
11180+1623 & 0.166 & 12.31$\pm$0.06 & 1.7 & $<3.5$ & 1.09 & 30407 & LINER & L & 1,15 \\
02113$-$2937 & 0.192 & 12.41$\pm$0.06 & $-$ & $<0.7$ & $-$ & 1096 & LINER & C & 5,19 & & 
11223$-$1244 & 0.199 & 12.57$\pm$0.04 & 5.6 & $<7.1$ & $<0.01$ & 105 & Sy 2 & L & 1,15 \\
02411+0354 & 0.144 & 12.27$\pm$0.04 & $-$ & $<0.9$ & $-$ & 2306 & H~\textsc{ii} & L & 1,15 & & 
11387+4116 & 0.149 & 12.21$\pm$0.10 & $-$ & $<0.04$ & $-$ & 2306 & H~\textsc{ii} & C & 1,15 \\
02456$-$2220 & 0.290 & 12.72$\pm$0.07 & 2.1 & 0.5$-$2.7 & $<0.09$ & 105 & $-$ & $-$ & $-$ & & 
11506+1331 & 0.127 & 12.36$\pm$0.04 & 4.5 & 1.7$-$5.2 & $<0.01$ & 3187 & H~\textsc{ii} & C & 12,17 \\
03000$-$2719 & 0.221 & 12.55$\pm$0.05 & 5.2 & 4.0$-$6.8 & 0.68 & 1096 & $-$ & C & 19 & & 
11524+1058 & 0.179 & 12.23$\pm$0.06 & 2.1 & $<4.1$ & $<0.54$ & 30407 & LINER & $-$ & 7 \\
03158+4227 & 0.134 & 12.61$\pm$0.04 & 47 & 42$-$52 & 1.72 & 105 & Sy 2 & L & 8 & & 
11553+4557 & 0.147 & 12.09$\pm$0.07 & 14 & 11$-$16 & 0.12 & 40991 & $-$ & $-$ & $-$ \\
03250+1606 & 0.129 & 12.12$\pm$0.06 & $-$ & $<0.2$ & $-$ & 3187 & LINER & C & 1,17 & & 
11582+3020 & 0.223 & 12.58$\pm$0.07 & 70 & 64$-$75 & 2.47 & 105 & LINER & C & 12,15 \\
03521+0028 & 0.152 & 12.60$\pm$0.06 & 0.8 & $<1.2$ & $<0.15$ & 105 & LINER & C & 1,15 & & 
12018+1941 & 0.169 & 12.52$\pm$0.05 & 43 & 39$-$48 & 1.69 & 105 & $-$ & C & 15 \\
03538$-$6432 & 0.301 & 12.79$\pm$0.04 & 14 & 12$-$16 & 0.36 & 105 & $-$ & C & 20 & & 
12032+1707 & 0.218 & 12.63$\pm$0.07 & 67 & 62$-$72 & 1.71 & 105 & LINER & L & 1,15 \\
04103$-$2838 & 0.117 & 12.23$\pm$0.02 & 5.4 & 4.6$-$6.2 & 0.14 & 3187 & LINER & C & 1,17 & & 
12071$-$0444 & 0.128 & 12.40$\pm$0.03 & 41 & 37$-$45 & 1.07 & 105 & Sy 2 & C & 12,17 \\
04114$-$5117 & 0.125 & 12.26$\pm$0.03 & $-$ & $<0.3$ & $-$ & 105 & $-$ & L & 16 & & 
12112+0305 & 0.073 & 12.33$\pm$0.02 & $-$ & $<0.7$ & $-$ & 105 & LINER & C & 12,22 \\
04313$-$1649 & 0.268 & 12.67$\pm$0.05 & 70 & 64$-$77 & 2.36 & 105 & $-$ & C & 17 & & 
12127$-$1412 & 0.133 & 12.20$\pm$0.06 & 88 & 85$-$91 & $<0.04$ & 3187 & LINER & L & 1,15 \\
04384$-$4848 & 0.203 & 12.41$\pm$0.03 & 4.9 & 3.9$-$6.1 & 0.62 & 1096 & $-$ & C & 18 & & 
12359$-$0725 & 0.138 & 12.18$\pm$0.10 & 4.5 & $<5.5$ & $<0.01$ & 2306 & LINER & L & 1,15 \\
04394$-$3740 & 0.237 & 12.56$\pm$0.05 & 21 & 18$-$25 & 0.73 & 3231 & Sy 2 & $-$ & 9 & & 
12514+1027 & 0.300 & 12.73$\pm$0.06 & 96 & $>79$ & 0.35 & 105 & $-$ & $-$ & $-$ \\
05024$-$1941 & 0.192 & 12.51$\pm$0.12 & 9.5 & 6.2$-$14 & 1.91 & 3187 & Sy 2 & C & 1,17 & & 
12540+5708 & 0.042 & 12.55$\pm$0.01 & 34 & 31$-$37 & $<0.12$ & 105 & Sy 1 & C & 12,15 \\
05189$-$2524 & 0.043 & 12.17$\pm$0.01 & 30 & 22$-$34 & $<0.01$ & 105 & Sy 2 & C & 10,17 & & 
13120$-$5453 & 0.031 & 12.25$\pm$0.03 & 0.8 & $<1.1$ & 0.12 & 105 & $-$ & $-$ & $-$ \\
06009$-$7716 & 0.117 & 12.05$\pm$0.03 & $-$ & $<0.3$ & $-$ & 666 & $-$ & L & 16 & & 
13218+0552 & 0.205 & 12.73$\pm$0.03 & 95 & $>44$ & $<0.01$ & 105 & Sy 1 & C & 12,15 \\
06035$-$7102 & 0.079 & 12.23$\pm$0.02 & 22 & 12$-$25 & $<0.01$ & 105 & LINER & L & 3 & & 
13335$-$2612 & 0.125 & 12.12$\pm$0.06 & $-$ & $<0.03$ & $-$ & 3187 & LINER & C & 1,15 \\
06206$-$6315 & 0.092 & 12.22$\pm$0.02 & 6.4 & 4.2$-$7.4 & $<0.01$ & 105 & Sy 2 & C & 3,16 & & 
13342+3932 & 0.179 & 12.40$\pm$0.08 & 18 & 7.3$-$21 & $<0.01$ & 105 & Sy 1 & C & 12,15 \\
06268+3509 & 0.170 & 12.34$\pm$0.14 & $-$ & $<0.3$ & $-$ & 20589 & H~\textsc{ii} & L & 7,20 & & 
13352+6402 & 0.234 & 12.54$\pm$0.04 & 12 & 7.6$-$14 & $<0.01$ & 105 & $-$ & L & 20 \\
06301$-$7934 & 0.156 & 12.39$\pm$0.03 & 9.6 & 7.4$-$12 & 1.74 & 105 & $-$ & C & 16 & & 
13428+5608 & 0.038 & 12.15$\pm$0.01 & 7.5 & 5.6$-$8.5 & $<0.01$ & 105 & Sy 2 & C & 12,22 \\
06361$-$6217 & 0.160 & 12.41$\pm$0.04 & 18 & 13$-$20 & $<0.01$ & 105 & $-$ & C & 20 & & 
13451+1232 & 0.122 & 12.31$\pm$0.04 & 60 & 54$-$65 & 0.24 & 105 & Sy 2 & L & 12,22 \\
06487+2208 & 0.143 & 12.44$\pm$0.09 & 2.6 & 2.0$-$3.3 & $<0.21$ & 30407 & H~\textsc{ii} & C & 7,21 & & 
13454$-$2956 & 0.129 & 12.30$\pm$0.03 & 5.8 & 1.8$-$6.8 & $<0.01$ & 3187 & Sy 2 & L & 12,15 \\
07246+6125 & 0.137 & 12.06$\pm$0.07 & 19 & 15$-$24 & $<0.06$ & 30121 & Sy 2 & C & 11,21 & & 
13469+5833 & 0.158 & 12.25$\pm$0.03 & 3.6 & 2.5$-$4.9 & $<0.20$ & 20589 & H~\textsc{ii} & C & 1,15 \\
07251$-$0248 & 0.088 & 12.41$\pm$0.04 & 39 & 35$-$43 & 1.37 & 30323 & $-$ & $-$ & $-$ & & 
13510+0442 & 0.136 & 12.29$\pm$0.04 & $-$ & $<0.03$ & $-$ & 2306 & H~\textsc{ii} & C & 1,15 \\
07572+0533 & 0.190 & 12.47$\pm$0.13 & 25 & 21$-$30 & 1.05 & 30407 & LINER & $-$ & 7 & & 
13539+2920 & 0.108 & 12.09$\pm$0.04 & $-$ & $<0.02$ & $-$ & 2306 & H~\textsc{ii} & L & 1,15 \\
07598+6508 & 0.148 & 12.54$\pm$0.02 & 85 & 81$-$88 & $<0.06$ & 105 & Sy 1 & C & 10,22 & & 
14026+4341 & 0.323 & 12.96$\pm$0.04 & 92 & 88$-$95 & $<0.05$ & 61 & Sy 1 & C & 13,24 \\
08201+2801 & 0.168 & 12.30$\pm$0.06 & 39 & 33$-$44 & 1.83 & 30407 & H~\textsc{ii} & C & 12,15 & & 
14060+2919 & 0.117 & 12.12$\pm$0.04 & $-$ & $<0.02$ & $-$ & 2306 & H~\textsc{ii} & C & 12,15 \\
08344+5105 & 0.097 & 12.01$\pm$0.03 & $-$ & $<0.6$ & $-$ & 40991 & $-$ & C & 21 & & 
14070+0525 & 0.264 & 12.83$\pm$0.05 & 30 & 26$-$35 & 1.40 & 105 & Sy 2 & C & 12,17 \\
08449+2332 & 0.151 & 12.14$\pm$0.09 & 3.2 & $<5.0$ & 1.22 & 30407 & H~\textsc{ii} & $-$ & 7 & & 
14197+0812 & 0.131 & 12.11$\pm$0.09 & 11 & 8.0$-$13 & 2.10 & 3187 & $-$ & C & 15\\
08559+1053 & 0.148 & 12.26$\pm$0.05 & 7.6 & 6.6$-$8.7 & $<0.01$ & 666 & Sy 2 & C & 1,15 & & 
14252$-$1550 & 0.150 & 12.23$\pm$0.07 & $-$ & $<1.1$ & $-$ & 2306 & LINER & L & 1,15 \\
08572+3915 & 0.058 & 12.15$\pm$0.01 & 86 & 83$-$88 & 0.44 & 105 & LINER & L & 12,22 & & 
14348$-$1447 & 0.083 & 12.35$\pm$0.02 & 3.9 & 2.8$-$5.0 & $<0.09$ & 105 & LINER & L & 12,22 \\
09022$-$3615 & 0.060 & 12.30$\pm$0.01 & 9.0 & 5.9$-$10 & $<0.01$ & 105 & $-$ & C & 23 & & 
14378$-$3651 & 0.068 & 12.13$\pm$0.02 & 1.0 & $<1.3$ & $<0.08$ & 105 & Sy 2 & C & 3 \\
\hline
\end{tabular}
\end{scriptsize}
\end{table*}
\end{landscape}

\clearpage
\thispagestyle{empty}
\begin{landscape}
\begin{table*}
\contcaption{}
\begin{scriptsize}
\begin{tabular}{ccccccccccccccccccccc}
\hline \hline
\\
Object & $z$ & $\log (L_\mathit{IR}/L_\odot)$ & $\alpha_\mathit{bol}$ & $\Delta \alpha$ & $\tau_6$ & PID 
& Type & IC & Refs. & &
Object & $z$ & $\log (L_\mathit{IR}/L_\odot)$ & $\alpha_\mathit{bol}$ & $\Delta \alpha$ & $\tau_6$ & PID 
& Type & IC & Refs. \\
(1) & (2) & (3) & (4) & (5) & (6) & (7) & (8) & (9) & (10) & & 
(1) & (2) & (3) & (4) & (5) & (6) & (7) & (8) & (9) & (10) \\
\\
\hline
15001+1433 & 0.163 & 12.46$\pm$0.04 & 7.9 & 2.9$-$9.5 & $<0.02$ & 105 & Sy 2 & L & 12,15 & & 
18588+3517 & 0.107 & 11.97$\pm$0.04 & 3.5 & 1.3$-$4.1 & $<0.01$ & 30407 &  H~\textsc{ii} & $-$ & 7 \\
15130$-$1958 & 0.109 & 12.16$\pm$0.05 & 30 & 21$-$33 & $<0.01$ & 3187 & Sy 2 & C & 12,17 & & 
19254$-$7245 & 0.062 & 12.10$\pm$0.01 & 24 & 20$-$28 & 0.21 & 105 & Sy 2 & L & 3 \\
15176+5216 & 0.139 & 12.15$\pm$0.02 & 37 & 26$-$46 & $<0.01$ & 40991 & Sy 2 & $-$ & 13 & & 
19297$-$0406 & 0.086 & 12.40$\pm$0.02 & 1.0 & $<1.3$ & $<0.08$ & 105 & H~\textsc{ii} & C & 10,20 \\
15206+3342 & 0.124 & 12.25$\pm$0.02 & 4.1 & 3.5$-$4.8 & 0.30 & 105 & H~\textsc{ii} & C & 12,15 & & 
19458+0944 & 0.100 & 12.37$\pm$0.05 & 0.8 & $<1.0$ & $<0.12$ & 105 & $-$ & C & 25 \\
15225+2350 & 0.139 & 12.18$\pm$0.04 & 21 & 18$-$23 & 0.74 & 2306 & H~\textsc{ii} & C & 1,15 & & 
19542+1110 & 0.065 & 12.12$\pm$0.04 & 3.8 & 3.0$-$4.5 & $<0.04$ & 30323 & $-$ & $-$ & $-$ \\
15250+3609 & 0.055 & 12.04$\pm$0.02 & 51 & 46$-$55 & 1.06 & 105 & LINER & C & 10,22 & & 
20037$-$1547 & 0.192 & 12.58$\pm$0.07 & 26 & 8.7$-$30 & $<0.02$ & 105 & Sy 1 & L & 11,20 \\
15327+2340 & 0.018 & 12.17$\pm$0.01 & 17 & 15$-$19 & 1.47 & 105 & LINER & C & 12,22 & & 
20087$-$0308 & 0.106 & 12.44$\pm$0.03 & 3.1 & $<3.8$ & $<0.01$ & 105 & LINER & C & 12,20 \\
15462$-$0450 & 0.100 & 12.21$\pm$0.04 & 26 & 8.9$-$31 & $<0.01$ & 105 & Sy 1 & C & 12,17 & & 
20100$-$4156 & 0.130 & 12.64$\pm$0.02 & 20 & 18$-$23 & 0.47 & 105 & H~\textsc{ii} & L & 3 \\
16090$-$0139 & 0.134 & 12.57$\pm$0.02 & 24 & 21$-$27 & 0.69 & 105 & LINER & C & 12,15 & & 
20286+1846 & 0.136 & 12.20$\pm$0.11 & $-$ & $<0.2$ & $-$ & 30407 & LINER & $-$ & 7 \\
16155+0146 & 0.132 & 12.11$\pm$0.06 & 40 & 36$-$44 & 0.61 & 3187 & Sy 2 & L & 1,15 & & 
20414$-$1651 & 0.087 & 12.30$\pm$0.08 & $-$ & $<0.1$ & $-$ & 105 & H~\textsc{ii} & C & 12,17 \\
16255+2801 & 0.134 & 12.04$\pm$0.05 & 23 & 19$-$28 & 0.63 & 30407 & H~\textsc{ii} & $-$ & 7 & & 
20551$-$4250 & 0.043 & 12.05$\pm$0.01 & 26 & 23$-$29 & 1.19 & 105 & LINER & C & 3 \\
16300+1558 & 0.242 & 12.74$\pm$0.05 & 9.4 & 7.9$-$11 & 0.30 & 105 & LINER & C & 12,17 & & 
21208$-$0519 & 0.130 & 12.07$\pm$0.06 & $-$ & $<0.04$ & $-$ & 3187 & H~\textsc{ii} & L & 1,15 \\
16334+4630 & 0.191 & 12.45$\pm$0.04 & 1.0 & $<1.6$ & $<0.01$ & 105 & LINER & L & 12,15 & & 
21219$-$1757 & 0.112 & 12.13$\pm$0.03 & 84 & 42$-$95 & $<0.01$ & 3187 & Sy 1 & C & 1,17 \\
16455+4553 & 0.191 & 12.37$\pm$0.04 & 7.7 & 5.9$-$9.9 & 0.43 & 20589 & $-$ & C & 20 & & 
21329$-$2345 & 0.125 & 12.15$\pm$0.04 & 2.9 & 2.4$-$3.5 & $<0.07$ & 3187 & LINER & C & 12,17 \\
16468+5200 & 0.150 & 12.11$\pm$0.05 & 19 & 16$-$21 & 0.77 & 2306 & LINER & L & 12,15 & & 
22055+3024 & 0.127 & 12.29$\pm$0.10 & 9.9 & 8.0$-$12 & 0.59 & 30407 & LINER & $-$ & 7 \\
16474+3430 & 0.111 & 12.20$\pm$0.04 & $-$ & $<0.3$ & $-$ & 2306 & H~\textsc{ii} & L & 12,15 & & 
22206$-$2715 & 0.131 & 12.23$\pm$0.05 & $-$ & $<0.5$ & $-$ & 3187 & H~\textsc{ii} & L & 1,15 \\
16487+5447 & 0.104 & 12.19$\pm$0.02 & 1.1 & $<1.3$ & $<0.04$ & 2306 & LINER & L & 12,15 & & 
22491$-$1808 & 0.078 & 12.19$\pm$0.03 & $-$ & $<0.07$ & $-$ & 105 & H~\textsc{ii} & C & 10,22 \\
16541+5301 & 0.194 & 12.25$\pm$0.06 & 6.0 & $<7.4$ & $<0.03$ & 20589 & Sy 2 & L & 11,20 & & 
23019+3405 & 0.108 & 11.99$\pm$0.04 & 0.8 & $<1.0$ & $<0.01$ & 30407 & Sy 2 & $-$ & 7 \\
17028+5817 & 0.106 & 12.17$\pm$0.02 & $-$ & $<0.06$ & $-$ & 2306 & LINER & L & 12,15 & & 
23060+0505 & 0.173 & 12.53$\pm$0.03 & 78 & 54$-$83 & $<0.01$ & 61 & Sy 2 & C & 1,15 \\
17044+6720 & 0.135 & 12.17$\pm$0.03 & 27 & 24$-$30 & 0.32 & 2306 & LINER & C & 12,15 & & 
23128$-$5919 & 0.045 & 12.04$\pm$0.01 & 3.6 & 3.0$-$4.2 & 0.36 & 105 & H~\textsc{ii} & C & 3 \\
17068+4027 & 0.179 & 12.40$\pm$0.04 & 15 & 13$-$17 & 0.87 & 105 & H~\textsc{ii} & L & 12,15 & & 
23129+2548 & 0.179 & 12.48$\pm$0.05 & 68 & 62$-$73 & 2.32 & 105 & LINER & C & 1,15 \\
17179+5444 & 0.147 & 12.28$\pm$0.04 & 17 & 15$-$19 & 0.31 & 105 & Sy 2 & C & 1,15 & & 
23230$-$6926 & 0.106 & 12.31$\pm$0.02 & 6.0 & 5.0$-$7.2 & 0.60 & 105 & LINER & C & 3 \\
17208$-$0014 & 0.043 & 12.40$\pm$0.02 & $-$ & $<0.4$ & $-$ & 105 & LINER & C & 3,22 & & 
23233+0946 & 0.128 & 12.15$\pm$0.05 & 1.7 & 1.3$-$2.0 & $<0.05$ & 3187 & LINER & L & 1,15 \\
17463+5806 & 0.309 & 12.64$\pm$0.05 & 4.8 & 3.8$-$6.1 & 0.12 & 105 & $-$ & C & 21 & & 
23253$-$5415 & 0.130 & 12.36$\pm$0.03 & 7.5 & 6.1$-$9.1 & 0.39 & 105 & H~\textsc{ii} & $-$ & 14 \\
18030+0705 & 0.146 & 12.38$\pm$0.09 & $-$ & $<0.2$ & $-$ & 105 & $-$ & $-$ & $-$ & & 
23327+2913 & 0.107 & 12.13$\pm$0.03 & 9.6 & 8.2$-$11 & 0.86 & 2306 & LINER & L & 1,15 \\
18368+3549 & 0.116 & 12.29$\pm$0.07 & $-$ & $<0.06$ & $-$ & 30407 & Sy 2 & C & 7,25 & & 
23365+3604 & 0.064 & 12.17$\pm$0.03 & 7.2 & 5.0$-$10 & 2.31 & 105 & LINER & C & 10,25 \\
18443+7433 & 0.135 & 12.33$\pm$0.03 & 23 & 20$-$26 & 0.79 & 105 & $-$ & C & 25 & & 
23498+2423 & 0.212 & 12.51$\pm$0.08 & 26 & 23$-$30 & $<0.01$ & 105 & Sy 2 & L & 1,15 \\
18531$-$4616 & 0.141 & 12.33$\pm$0.09 & 17 & 13$-$23 & 3.39 & 105 & $-$ & $-$ & $-$ & & 
23515$-$2917 & 0.335 & 12.81$\pm$0.08 & 23 & 20$-$27 & $<0.01$ & 30073 & Sy 2 & $-$ & 5 \\
18580+6527 & 0.176 & 12.21$\pm$0.04 & 1.4 & $<1.8$ & $<0.04$ & 20589 & Sy 2 & L & 11,20 & & 
23578$-$5307 & 0.125 & 12.21$\pm$0.04 & 0.8 & $<1.5$ & $<0.41$ & 30202 & $-$ & C & 16 \\
\hline
\end{tabular}
\end{scriptsize}
\begin{flushleft}
\textit{References:} $^1$Veilleux et al.~(1999), $^2$Armus et al.~(1989), 
$^3$Duc, Mirabel \& Maza~(1997), $^4$Vader \& Simon~(1987), $^5$Allen et al.~(1991), 
$^6$Kewley et al.~(2001), $^7$Darling \& Giovanelli~(2006), $^8$Meusinger et al.~(2001), 
$^9$Leipski et al.~(2007), $^{10}$Veilleux et al.~(1995), $^{11}$Lawrence et al~(1999), 
$^{12}$Kim et al.~(1998), $^{13}$de Grijp et al.~(1992), $^{14}$Sekiguchi \& Wolstencroft~(1993), 
$^{15}$Kim et al.~(2002), $^{16}$Rigopoulou et al.~(1999), $^{17}$Veilleux et al.~(2006), 
$^{18}$Farrah et al.~(2001), $^{19}$Clements et al.~(1996), $^{20}$Bushouse et al.~(2002), 
$^{21}$Cui et al.~(2001), $^{22}$Scoville et al.~(2000), $^{23}$Arribas et al.~(2008), 
$^{24}$Hutchings \& Morris~(1995), $^{25}$Murphy et al.~(1996). \\
\textit{Programmes (PIs):} 61~-~Intrinsic spectra of hyperluminous IR galaxies (G.Rieke); 
105~-~Spectroscopic study of distant ULIRGs II (J.R.Houck); 666~-~IRS campaign~P (J.R.Houck); 
1096~-~Dust in ULIRG environments (H.Spoon); 2306~-~Buried AGN in ultraluminous IR galaxies 
(M.Imanishi); 3187~-~The evolution of activity in massive gas-rich mergers (S.Veilleux); 
3231~-~Establishing the mid-IR selection of AGN (M.Haas); 20589~-~The role of mergers and 
interactions in luminous and ultraluminous IR galaxies (C.Leitherer); 30073~-~IRS GTO ULIRG 
program: filling in the gaps (J.R.Houck); 30121~-~IRS spectra of extreme \textit{IRAS} sources 
(J.R.Houck); 30202~-~Revealing the nature of one of the most H$_2$-bright ULIRGs (H.Dannerbauer); 
30323~-~A \textit{Spitzer} spectroscopic survey of a complete sample of luminous IR galaxies in 
the local Universe (L.Armus); 30407~-~The astrophysics of OH megamasers in merging galaxies: the 
role of star formation, dust, molecules, and AGN (J.Darling); 40991~-~An IRS survey of IR sources 
in the \textit{IRAS} and SDSS (L.Hao). \\
\textit{Notes:} a) Whenever the confidence range is provided as an upper limit, the AGN detection is not 
considered as confident, with the exceptions of 00482$-$2720, 09463+8141, 11223$-$1244, 
12359$-$0725, 16541+5301 and 20087$-$0308. b) In the following cases the best fit has been obtained 
by allowing the AGN intrinsic slope to assume the value shown in brackets, instead of the template 1.5: 
00182$-$7112 (0.43), 00188$-$0856 (0.78), 01572+0009 (0.96), 07598+6508 (0.58), 09320+6134 (1.02), 
11119+3257 (0.47), 12127$-$1412 (1.31), 12514+1027 (1.00), 14026+4341 (1.17), 19254$-$7245 (0.52), 
20037$-$1547 (0.79), 23498+2423 (0.31), 23515$-$2917 (0.42). c) Comparable minima in the parameter 
space suggest the presence of a faint AGN in 02411+0354, while 14197+0812, 18531$-$4616 and 
23365+3604 can be SB-dominated. 
\end{flushleft}
\end{table*}
\end{landscape}

\begin{table*}
\caption{Summary of AGN contribution.}
\label{t2}
\begin{tabular}{lcclc}
\hline \hline
Class & $\alpha_\mathit{agn}$ & & $\log (L_\mathit{IR}/L_\odot)$ & $\alpha_\mathit{agn}$ \\
\hline
\vspace{5pt}
H~\textsc{ii} regions & 18.2$^{+1.4}_{-1.4}$ & & 12.0$-$12.3 & 13.6$^{+0.4}_{-0.6}$ \\
\vspace{5pt}
LINERs & 26.5$^{+1.1}_{-1.1}$ & & 12.3$-$12.5 & 14.8$^{+0.6}_{-0.7}$ \\
\vspace{5pt}
Seyfert 2s & 27.7$^{+1.4}_{-1.8}$ & & 12.5$-$12.7 & 37.1$^{+1.5}_{-2.1}$ \\
\vspace{5pt}
Seyfert 1s & 61.9$^{+2.8}_{-7.3}$ & & 12.7$-$13.0 & 56.6$^{+2.6}_{-4.6}$ \\
Unclassified & 18.6$^{+1.1}_{-1.5}$ & & Total & 27.4$^{+0.6}_{-1.0}$ \\
\hline
\end{tabular}
\end{table*}

\begin{table*}
\caption{Properties of the 52 high-redshift sources in the control sample. $z$, $D_L$: 
redshift and luminosity distance (in Gpc). $\alpha_\mathit{bol}$, $\Delta \alpha$: 
AGN bolometric contribution~(in per cent) and 1$\sigma$ confidence range. $\tau_6$: 
AGN optical depth. $f_\mathit{obs}$: \textit{IRAS} 60~$\mu$m or \textit{Spitzer}-MIPS 70~$\mu$m
flux density (in mJy). $L_\mathit{IR}$ (1): IR luminosity (in log of solar units) computed as 
$L_\mathit{IR}=4\pi (D_L)^2 \times 29.3 (f_{30\mu m})^{0.93}$, as detailed in the 
text. $L_\mathit{IR}$ (2): tabulated IR luminosity, converted to the \textit{WMAP} cosmology. 
PID: \textit{Spitzer} observational programme. Ref.: reference for the tabulated IR luminosity.}
\label{t3}
\begin{scriptsize}
\begin{tabular}{lcccccccccc}
\hline \hline
Object & $z$ & $D_L$ & $\alpha_\mathit{bol}$ & $\Delta \alpha$ & $\tau_6$ & $f_\mathit{obs}$ & 
$L_\mathit{IR}$ (1) & $L_\mathit{IR}$ (2) & PID & Ref. \\
\hline
IRAS F00235+1024 & 0.58 & 3.37 & 62 & 56$-$68 & 1.99 & 428$\pm$56 & 
13.17$\pm$0.23 & 13.15$\pm$0.12 & 3746 & 1 \\
IRAS F00476$-$0054 & 0.73 & 4.50 & 63 & 56$-$70 & 0.73 & 260$\pm$52 & 
13.30$\pm$0.23 & $>12.85$ & 105 & 2 \\
IRAS Z01368+0100 & 0.61 & 3.64 & 74 & 65$-$84 & 0.56 & 230$\pm$46 & 
13.00$\pm$0.25 & $>12.64$ & 105 & 3 \\
IRAS Z02433+0110 & 0.80 & 5.05 & 82 & 78$-$86 & 1.31 & 210$\pm$42 & 
13.36$\pm$0.21 & $>12.99$ & 105 & 3 \\
IRAS F10026+4949$^{a}$ & 1.12 & 7.70 & 99 & $>96$ & $-$ & 266$\pm$40 & 
13.98$\pm$0.18 & 14.01$\pm$0.08 & 82 & 1 \\
IRAS F10398+3247 & 0.63 & 3.79 & 31 & 27$-$39 & $<0.02$ & 189$\pm$36 & 
12.97$\pm$0.24 & $>12.76$ & 105 & 2 \\
IRAS F12509+3122 & 0.78 & 4.91 & 98 & $>53$ & $<0.01$ & 218$\pm$44 & 
13.34$\pm$0.22 & 13.41$\pm$0.05 & 3746 & 1 \\
IRAS F14481+4454 & 0.66 & 3.99 & 79 & 63$-$84 & $<0.01$ & 190$\pm$32 & 
13.04$\pm$0.23 & 12.91$\pm$0.12 & 30121 & 4 \\
IRAS F14503+6006 & 0.58 & 3.39 & 54 & 48$-$59 & 0.62 & 226$\pm$25 & 
12.91$\pm$0.22 & 12.54$\pm$0.20 & 30121 & 4 \\
IRAS F14537+1950 & 0.66 & 3.96 & 6.4 & 4.4$-$9.2 & 0.26 & 283$\pm$42 & 
13.19$\pm$0.22 & $<13.55$ & 105 & 5 \\
IRAS F15307+3252 & 0.93 & 6.08 & 47 & 41$-$53 & 0.18 & 234$\pm$35 & 
13.63$\pm$0.17 & $>13.13$ & 105 & 2 \\
IRAS F16001+1652 & 0.67 & 4.08 & 99 & $>89$ & 0.27 & 153$\pm$41 & 
12.97$\pm$0.27 & 12.94$\pm$0.12 & 30121 & 4 \\
IRAS F16124+3241 & 0.71 & 4.37 & 29 & 23$-$36 & 1.29 & 174$\pm$33 & 
13.11$\pm$0.23 & 13.16$\pm$0.07 & 105 & 1 \\
IRAS F17233+3712 & 0.69 & 4.21 & 74 & 68$-$80 & 1.89 & 196$\pm$29 & 
13.11$\pm$0.22 & $>12.79$ & 105 & 3 \\
IRAS Z21293$-$0154 & 0.73 & 4.52 & 45 &3 8$-$53 & 1.73 & 190$\pm$38 & 
13.18$\pm$0.23 & $>12.83$ & 105 & 2 \\
\hline
ELAISC15 J003640$-$433925$^{b}$ & 1.18 & 8.23 & 82 & $>65$ & $-$ & 33.1$\pm$1.7 & 
13.01$\pm$0.15 & 12.5$\pm$0.5 & 3640 & 6 \\
ELAISC15 J004055$-$441249 & 1.38 & 10.0 & 68 & 27$-$83 & $<0.01$ & 35.3$\pm$1.2 & 
13.29$\pm$0.13 & 12.9$\pm$0.1 & 3640 & 6 \\
ELAISC15 J160733.7+534749 & 0.61 & 3.62 & $-$ & $<0.4$ & $-$ & 36.1$\pm$1.3 & 
12.04$\pm$0.25 & 11.9$\pm$0.2 & 3640 & 6 \\
ELAISC15 J161015.6+540615 & 1.02 & 6.84 & 23 & 17$-$32 & 0.23 & 55.2$\pm$1.6 & 
12.98$\pm$0.17 & 12.5$\pm$0.1 & 3640 & 6 \\
ELAISC15 J161255.1+540724 & 0.91 & 5.92 & 12 & 0.8$-$16 & $<0.02$ & 21.8$\pm$1.1 & 
12.43$\pm$0.20 & 12.1$\pm$0.2 & 3640 & 6 \\
ELAISC15 J161350.0+542631 & 1.15 & 8.00 & $-$ & $<0.3$ & $-$ & 16.4$\pm$1.0 & 
12.69$\pm$0.16 & 12.4$\pm$0.1 & 3640 & 6 \\
ELAISC15 J161441.1+550208 & 1.29 & 9.22 & 54 & 45$-$64 & 0.11 & 24.5$\pm$0.8 & 
13.04$\pm$0.13 & 12.8$\pm$0.1 & 3640 & 6 \\
ELAISC15 J161551.4+550722 & 1.10 & 7.53 & 15 & 2.6$-$19 & $<0.02$ & 27.5$\pm$1.2 & 
12.82$\pm$0.16 & 12.4$\pm$0.1 & 3640 & 6 \\
ELAISC15 J163422.0+414350 & 1.03 & 6.92 & 21 & 15$-$28 & 0.34 & 26.6$\pm$0.9 & 
12.70$\pm$0.17 & 12.5$\pm$0.2 & 3640 & 6 \\
ELAISC15 J163531.1+410025$^{c}$ & 1.15 & 7.98 & 61 & 48$-$77 & $-$ & 15.2$\pm$1.1 & 
12.66$\pm$0.17 & 12.4$\pm$0.5 & 3640 & 6 \\
ELAISC15 J163536.6+404754 & 0.62 & 3.69 & $-$ & $<0.9$ & $-$ & 22.7$\pm$1.4 & 
11.88$\pm$0.26 & 11.9$\pm$0.1 & 3640 & 6 \\
ELAISC15 J163553.5+412054 & 1.20 & 8.35 & 99 & $>93$ & 0.17 & 19.9$\pm$1.0 & 
12.82$\pm$0.15 & 12.4$\pm$0.1 & 3640 & 6 \\
ELAISC15 J163739.2+405643 & 1.43 & 10.4 & 25 & 20$-$31 & 0.10 & 27.5$\pm$1.0 & 
13.24$\pm$0.14 & 12.8$\pm$0.1 & 3640 & 6 \\
ELAISC15 J164010.1+410521$^{d}$ & 1.10 & 7.52 & 87 & 79$-$95 & $-$ & 44.3$\pm$1.2 & 
13.01$\pm$0.15 & 12.5$\pm$0.1 & 3640 & 6 \\
ELAISC15 J164036.8+412524 & 1.20 & 8.37 & 76 & 66$-$85 & 1.11 & 15.0$\pm$1.2 & 
12.71$\pm$0.16 & 12.5$\pm$0.2 & 3640 & 6 \\
SST24 J142651.9+343135 & 0.50 & 2.85 & $-$ & $<1.0$ & $-$ & 32.8$\pm$4.2 & 
11.73$\pm$0.31 & 11.94$\pm$0.22 & 15 & 7 \\
SST24 J143050.8+344848 & 1.21 & 8.49 & 97 & $>82$ & 0.98 & 43.2$\pm$5.1 & 
13.16$\pm$0.17 & 13.25$\pm$0.06 & 15 & 7 \\
SST24 J143151.8+324327 & 0.66 & 4.02 & $-$ & $<2.2$ & $-$ & 51.8$\pm$4.7 & 
12.31$\pm$0.26 & 12.51$\pm$0.11 & 15 & 7 \\
SST24 J143218.1+341300 & 0.98 & 6.48 & $-$ & $<1.5$ & $-$ & 36.6$\pm$5.1 & 
12.75$\pm$0.22 & 12.73$\pm$0.22 & 15 & 7 \\
SST24 J143341.9+330136 & 0.81 & 5.14 & 70 & 53$-$85 & 1.86 & 63.2$\pm$3.9 & 
12.69$\pm$0.21 & 12.89$\pm$0.05 & 15 & 7 \\
SST24 J143449.3+341014$^{e}$ & 0.51 & 2.93 & 18 & 14$-$22 & $-$ & 94.5$\pm$4.5 & 
12.19$\pm$0.27 & 12.21$\pm$0.22 & 15 & 7 \\
SST24 J143639.0+345222 & 0.99 & 6.57 & $-$ & $<0.4$ & $-$ & 35.0$\pm$6.3 & 
12.75$\pm$0.23 & 12.81$\pm$0.25 & 15 & 7 \\
SST24 J143820.7+340233 & 0.67 & 4.05 & $-$ & $<0.3$ & $-$ & 67.2$\pm$3.1 & 
12.43$\pm$0.24 & 12.72$\pm$0.10 & 15 & 7 \\
SST24 J143830.6+344412 & 0.94 & 6.19 & 36 & 13$-$43 & $<0.03$ & 45.2$\pm$4.0 & 
12.78$\pm$0.20 & $<13.02$ & 15 & 7 \\
SWIRE4 J103637.18+584217.0 & 0.97 & 6.44 & 62 & 50$-$74 & 1.74 & 34.7$\pm$6.9 & 
12.72$\pm$0.24 & 12.42$\pm$0.11 & 30364 & 8 \\
SWIRE4 J103946.28+582750.7 & 0.90 & 5.86 & 27 & 7.2$-$58 & 1.57 & 22.8$\pm$4.6 & 
12.43$\pm$0.26 & 12.24$\pm$0.11 & 30364 & 8 \\
SWIRE4 J104057.84+565238.9 & 0.93 & 6.11 & $-$ & $<1.8$ & $-$ & 24.2$\pm$4.8 & 
12.51$\pm$0.25 & 12.29$\pm$0.11 & 30364 & 8 \\
SWIRE4 J104117.93+595822.9 & 0.65 & 3.92 & 1.9 & $<6.6$ & $<1.26$ & 32.9$\pm$6.6 & 
12.10$\pm$0.30 & 12.07$\pm$0.11 & 30364 & 8 \\
SWIRE4 J104439.45+582958.5 & 0.68 & 4.14 & 3.8 & $<6.9$ & $<0.36$ & 22.0$\pm$4.4 & 
12.00$\pm$0.30 & 12.03$\pm$0.11 & 30364 & 8 \\
SWIRE4 J104830.58+591810.2 & 0.94 & 6.19 & $-$ & $<0.4$ & $-$ & 20.6$\pm$4.1 & 
12.46$\pm$0.25 & 12.56$\pm$0.11 & 30364 & 8 \\
SWIRE4 J105432.71+575245.6 & 1.02 & 6.85 & 1.5 & $<2.7$ & $<0.37$ & 37.0$\pm$7.4 & 
12.82$\pm$0.24 & 12.51$\pm$0.11 & 30364 & 8 \\
SWIRE4 J105509.00+584934.3 & 0.88 & 5.70 & 3.7 & $<7.4$ & $<0.40$ & 24.1$\pm$4.8 & 
12.42$\pm$0.26 & 12.36$\pm$0.11 & 30364 & 8 \\
SWIRE4 J105840.62+582124.7 & 0.89 & 5.78 & $-$ & $<0.5$ & $-$ & 19.3$\pm$3.9 & 
12.35$\pm$0.26 & 12.37$\pm$0.11 & 30364 & 8 \\
SWIRE4 J105943.83+572524.9 & 0.80 & 5.06 & 9.0 & 3.5$-$13 & 0.39 & 30.7$\pm$6.1 & 
12.38$\pm$0.27 & 12.49$\pm$0.11 & 30364 & 8 \\
SMM J123635.5+621238 & 1.23 & 8.66 & $-$ & $<0.2$ & $-$ & 13.9$\pm$1.8 & 
12.73$\pm$0.18 & 12.6$\pm$0.1 & 20456 & 9 \\
SST24 J142552.71+340240.2 & 0.56 & 3.29 & 8.3 & 6.1$-$11 & 0.46 & 215$\pm$43 & 
12.65$\pm$0.31 & 12.58$\pm$0.20 & 20113 & 10 \\
MIPS J142824.0+352619 & 1.33 & 9.50 & $-$ & $<0.8$ & $-$ & 34$\pm$6 & 
13.21$\pm$0.18 & 13.51$\pm$0.09 & 15 & 11 \\
\hline
\end{tabular}
\end{scriptsize}
\begin{flushleft}
\textit{References.} $^1$Farrah et al.~(2002), $^2$Yang et al.~(2007), $^3$Stanford et al.~(2000), 
$^4$Sargsyan et al.~(2008), $^5$This work, by using equation~(\ref{e1}), 
$^6$Hern{\'a}n-Caballero et al.~(2009), $^7$Brand et al.~(2008), $^8$Farrah et al.~(2009a), 
$^9$Pope et al.~(2008), $^{10}$Houck et al.~(2007), $^{11}$Desai et al.~(2006). \\
\textit{Programmes (PIs):} 15~-~Seeking redshifts for optically unidentifiable IR 
sources (J.R.Houck); 82~-~The far-IR SEDs of luminous AGN (G.Rieke); 105~-~See 
Table~\ref{t1}; 3640~-~IRS observations of ultraluminous ELAIS galaxies (I.P{\'e}rez-Fournon); 
3746~-~A spectroscopic study of local hyperluminous IR galaxies (A.Verma); 
20113~-~Probing the moderate redshift galaxies mid- and far-IR SED (H.Dole); 
20456~-~Balancing the cosmic energy budget between AGN and SBs in the GOODS (R.Chary); 
30121~-~See Table~\ref{t1}; 30364~-~A systematic \textit{Spitzer}-IRS survey of obscured 
SB galaxies at $1.0<z<1.9$ (J.R.Houck). \\
\textit{Notes:} The best fits are obtained by allowing for the following values of the AGN intrinsic 
slope: $^a$1.36, $^b$1.18, $^c$0.98, $^d$1.10, $^e$0.70.
\end{flushleft}
\end{table*}


\appendix
\section{Additional notes}

1) As mentioned in Section~2, a few sources with available \textit{Spitzer}-IRS observations have been rejected 
from the final sample because not detected at 5--8~$\mu$m. In all these five cases, no meaningful spectra 
can be extracted from the entire SL spectral orders, while the detection is indeed solid in the LL ones. 
For IRAS~06561+1902, IRAS~08311$-$2459 and IRAS~20176$-$4756, the flux density above 
$\sim$15~$\mu$m is not compatible with a smooth connection to the SL orders: this suggests possible 
pointing errors rather than insufficient exposures. In fact, IRAS~08311$-$2459 is a very bright source 
whose spectral features are actually evident at the shorter wavelengths as well; however, the required SL1 to 
LL2 scaling factor is $>4$ and can not be safely attributed to aperture losses. Hence also this source has been 
rejected. The other two objects (IRAS~05233$-$2334 and IRAS~07381+3215) do not show clear evidence 
of flux discontinuity, since they are unusually weak also in the LL2 order. In principle, we can not rule out the 
possibility that these sources are really too faint at 5--8~$\mu$m. As already pointed out in Paper~II, it is unlikely that 
these are extremely \textit{red} AGN, completely obscured up to $\sim$15--20~$\mu$m. If this is the case, such objects 
are confirmed to be very rare, representing $\sim$1 per cent at most of the local ULIRG population. \\
2) Concerning the sources of the 1~Jy ULIRG sample, 93 out of 118 have been included in our sample. Apart 
from 3C273, eight of the missing sources have not been observed by \textit{Spitzer}, while the other 16 were not 
publicly available for the present analysis. These are IRAS~02480$-$3745, 03209$-$0806, 04074$-$2801, 05020$-$2941, 
08474+1813, 08592+5248, 10594+3818, 12447+3721, 13106$-$0922, 14121$-$0126, 14202+2615, 14394+5332, 
14484$-$1958, 15043+5754, 21477+0502, 22088$-$1831. All but IRAS~14394+5332 have been recently discussed 
in Imanishi et al. (2010). None of these sources is optically classified as a Seyfert, as they are evenly distributed 
among the unclassified, LINER and H~\textsc{ii} classes. Anyway, this does not introduce a significant bias in our 
sample, since the fraction of type 1 Seyferts is very low among ULIRGs, and the mid-IR spectral properties as well 
as the average AGN content within all the other optical types are remarkably similar, as proved in Section~5. \\
3) With just few exceptions, only the \textit{Spitzer}-IRS spectra of the 1-Jy sources have already been presented 
so far. A wide discussion of the low- and/or high-resolution 5--35~$\mu$m spectra can be found in Armus et al. (2007), 
Farrah et al. (2007), Imanishi et al. (2007), Imanishi (2009) and Veilleux et al. (2009a). The 5--8~$\mu$m spectra and  
the results of our AGN/SB decomposition for each of the 164 sources in our local sample are provided as online material: 
following the same code of Paper~II, the SB component (\textit{red dot-dashed line}) and the observed AGN continuum 
(\textit{blue long-dashed line}) are shown alongside the data points (\textit{green filled circles}) and the best fit model 
(\textit{black solid line}), as in Fig.~\ref{apf}. 

\begin{figure*}
\includegraphics[width=\linewidth]{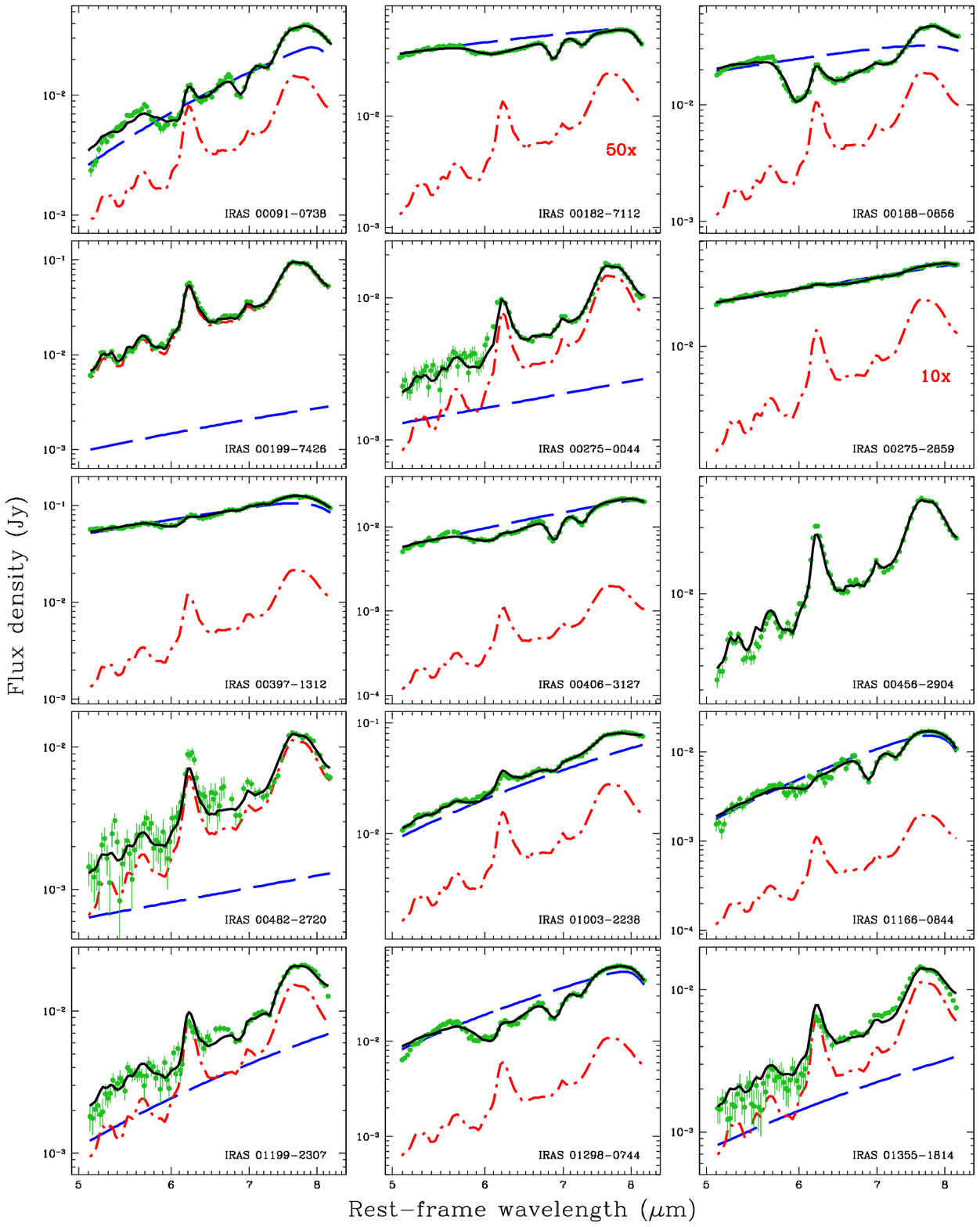}
\caption{Best fits and spectral components. The remaining panels (2--11) are available in the online version of the article.} 
\label{apf}
\end{figure*}
\begin{figure*}
\includegraphics[width=\linewidth]{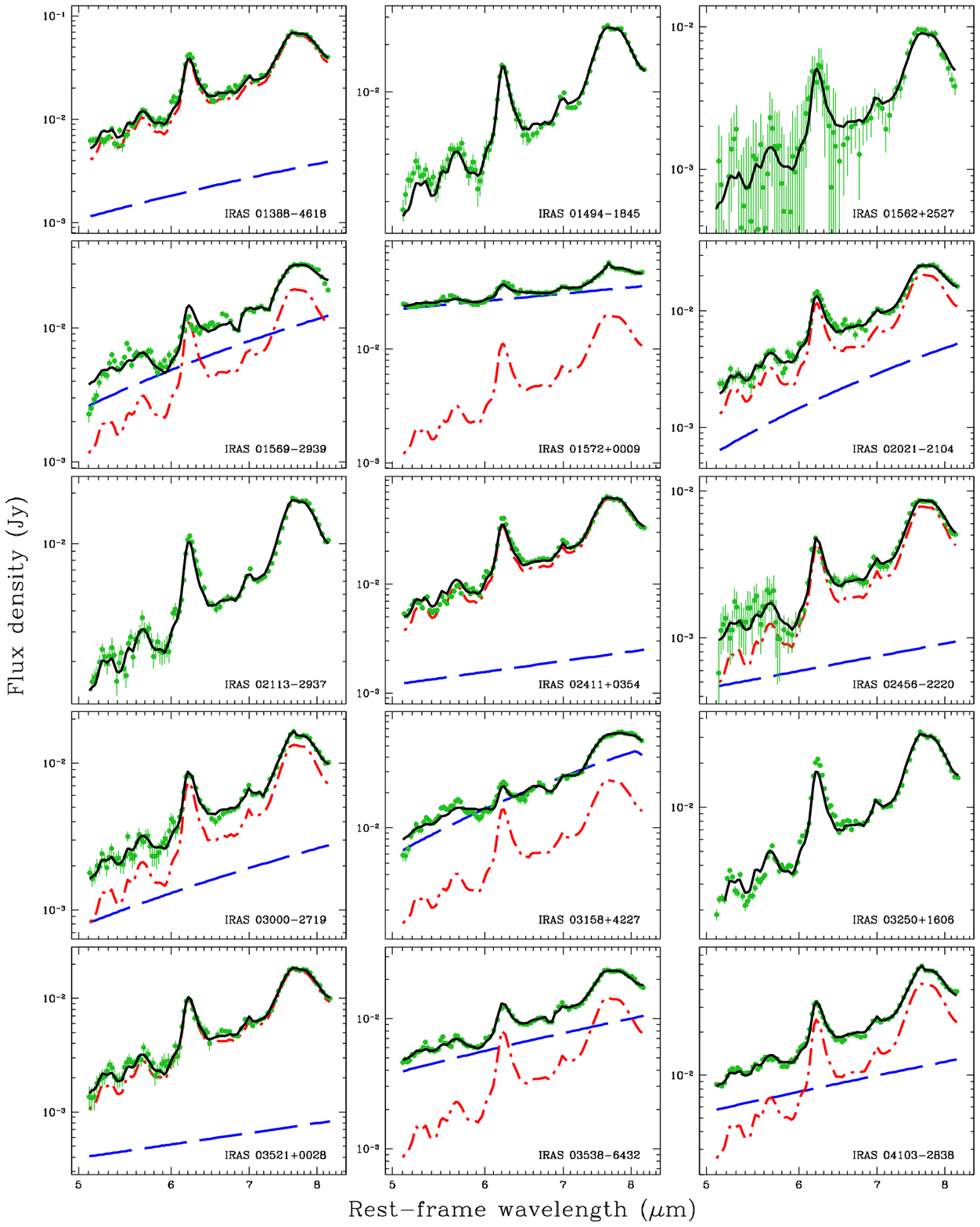}
\contcaption{} 
\end{figure*}
\begin{figure*}
\includegraphics[width=\linewidth]{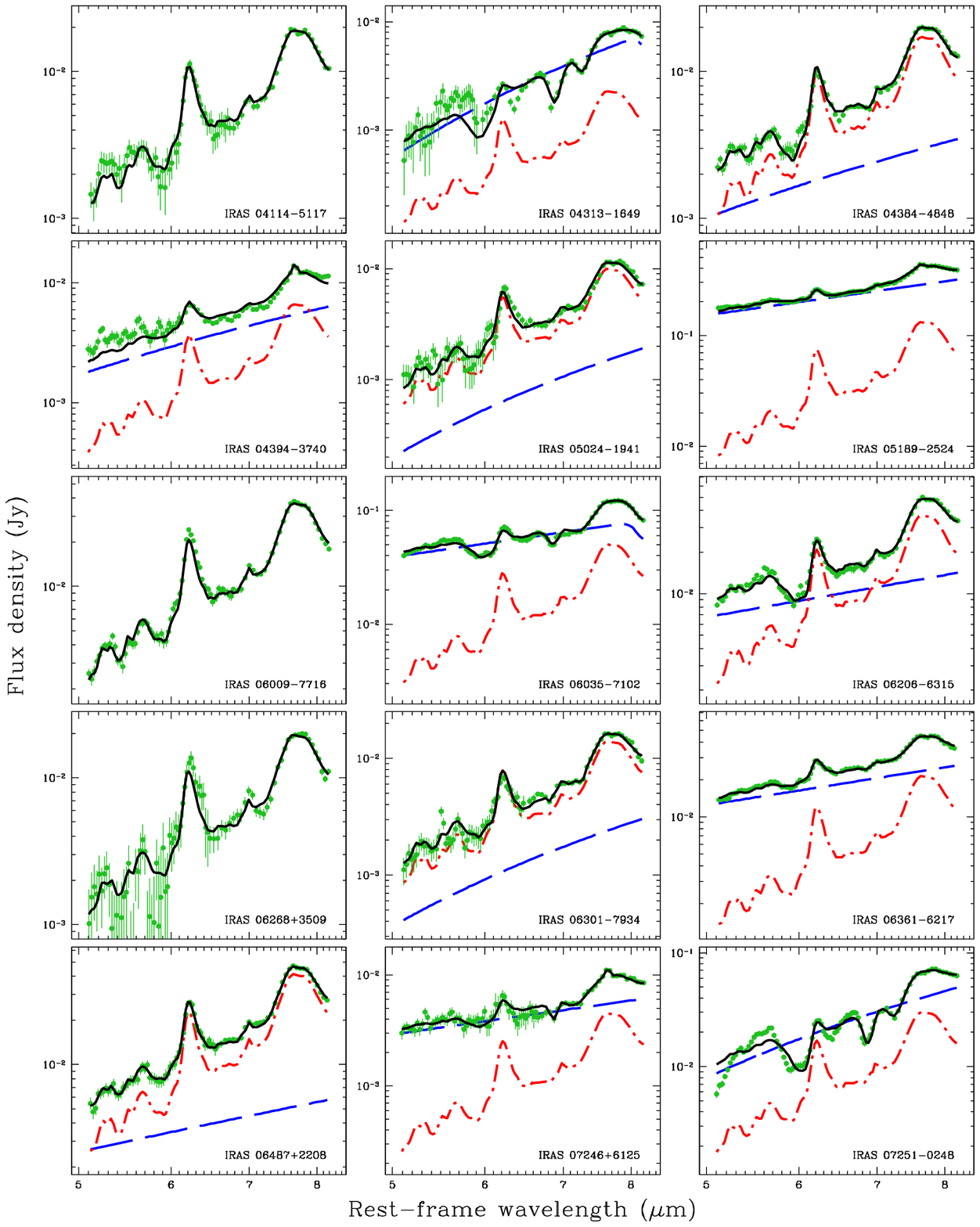}
\contcaption{} 
\end{figure*}
\begin{figure*}
\includegraphics[width=\linewidth]{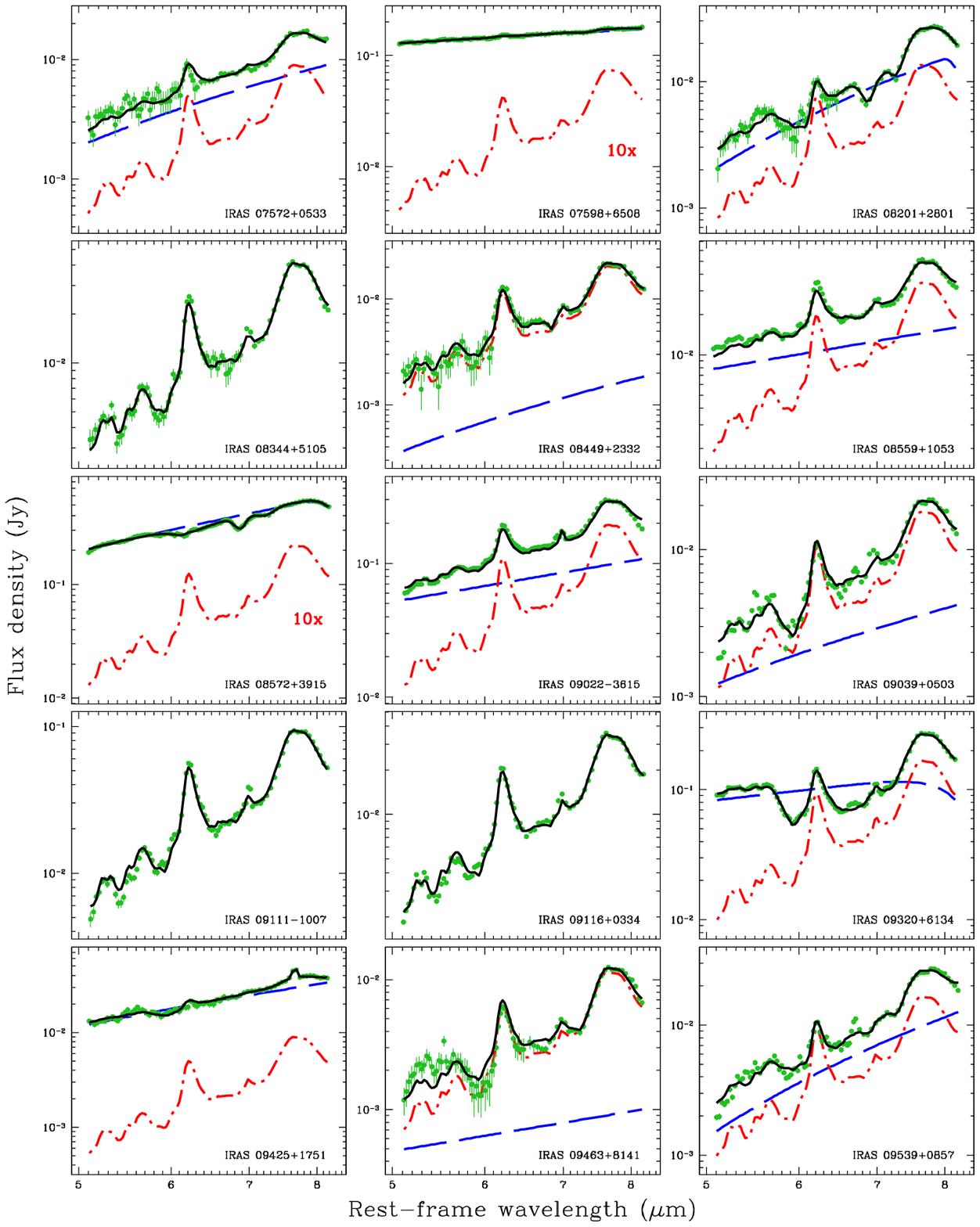}
\contcaption{} 
\end{figure*}
\begin{figure*}
\includegraphics[width=\linewidth]{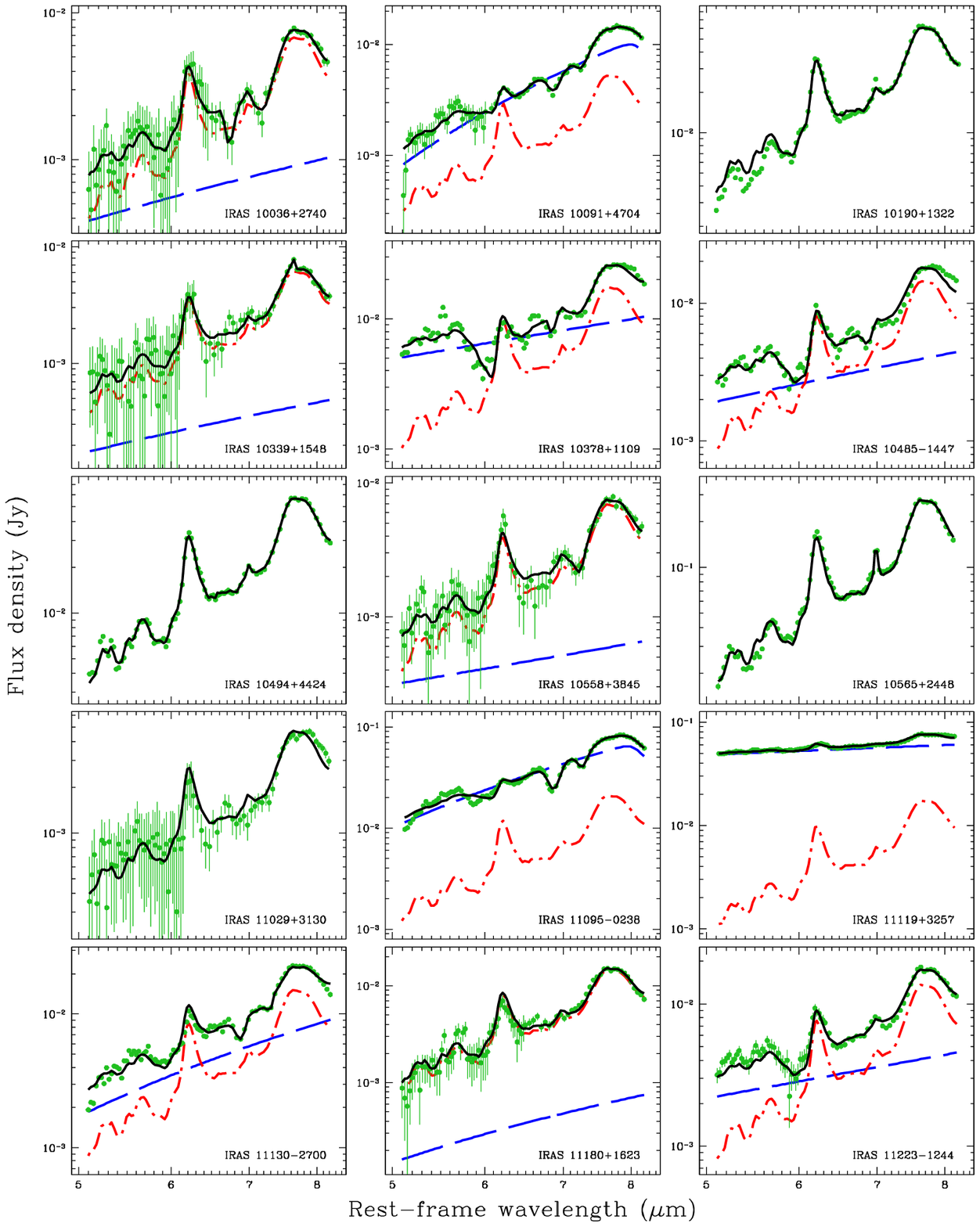}
\contcaption{} 
\end{figure*}
\begin{figure*}
\includegraphics[width=\linewidth]{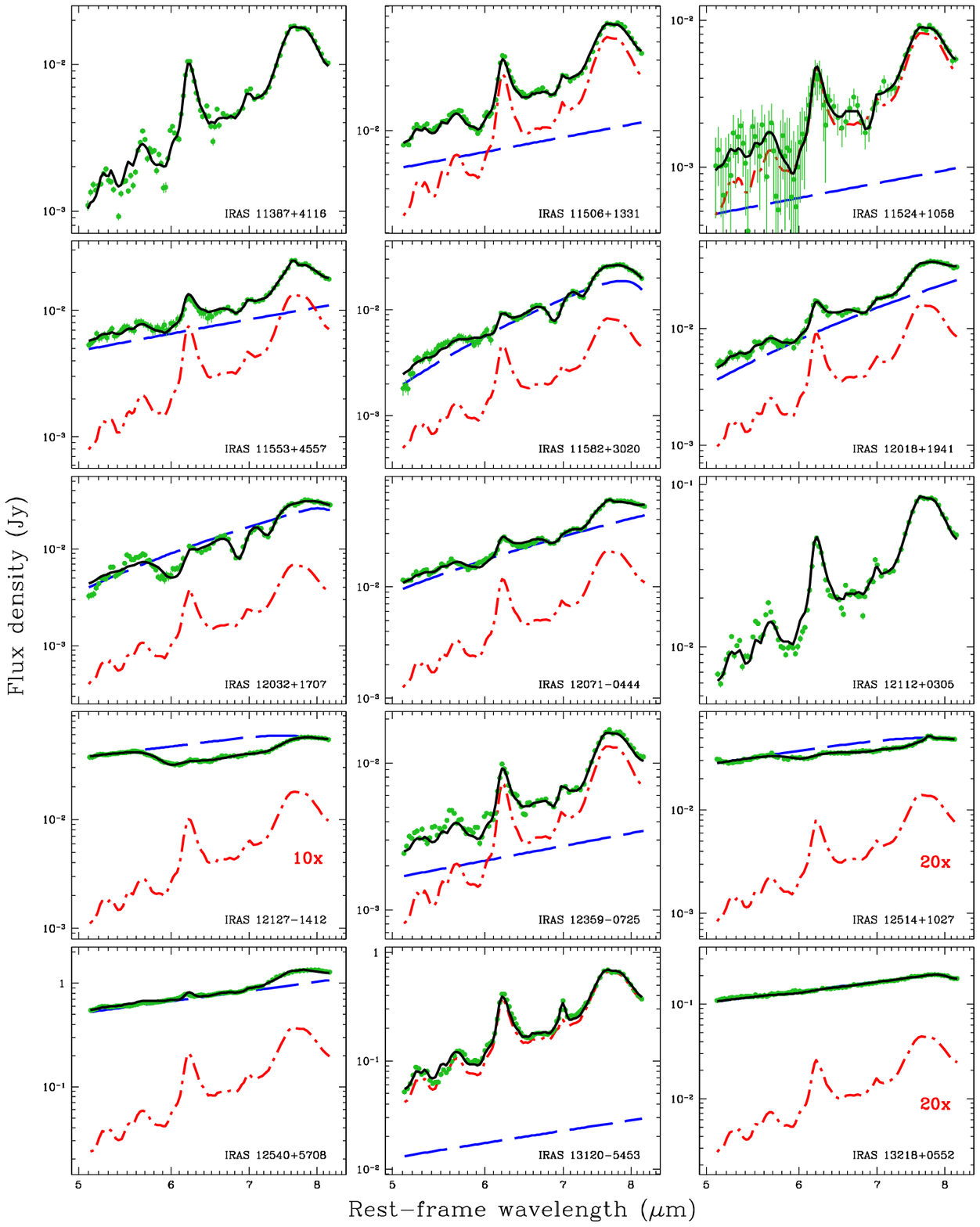}
\contcaption{} 
\end{figure*}
\begin{figure*}
\includegraphics[width=\linewidth]{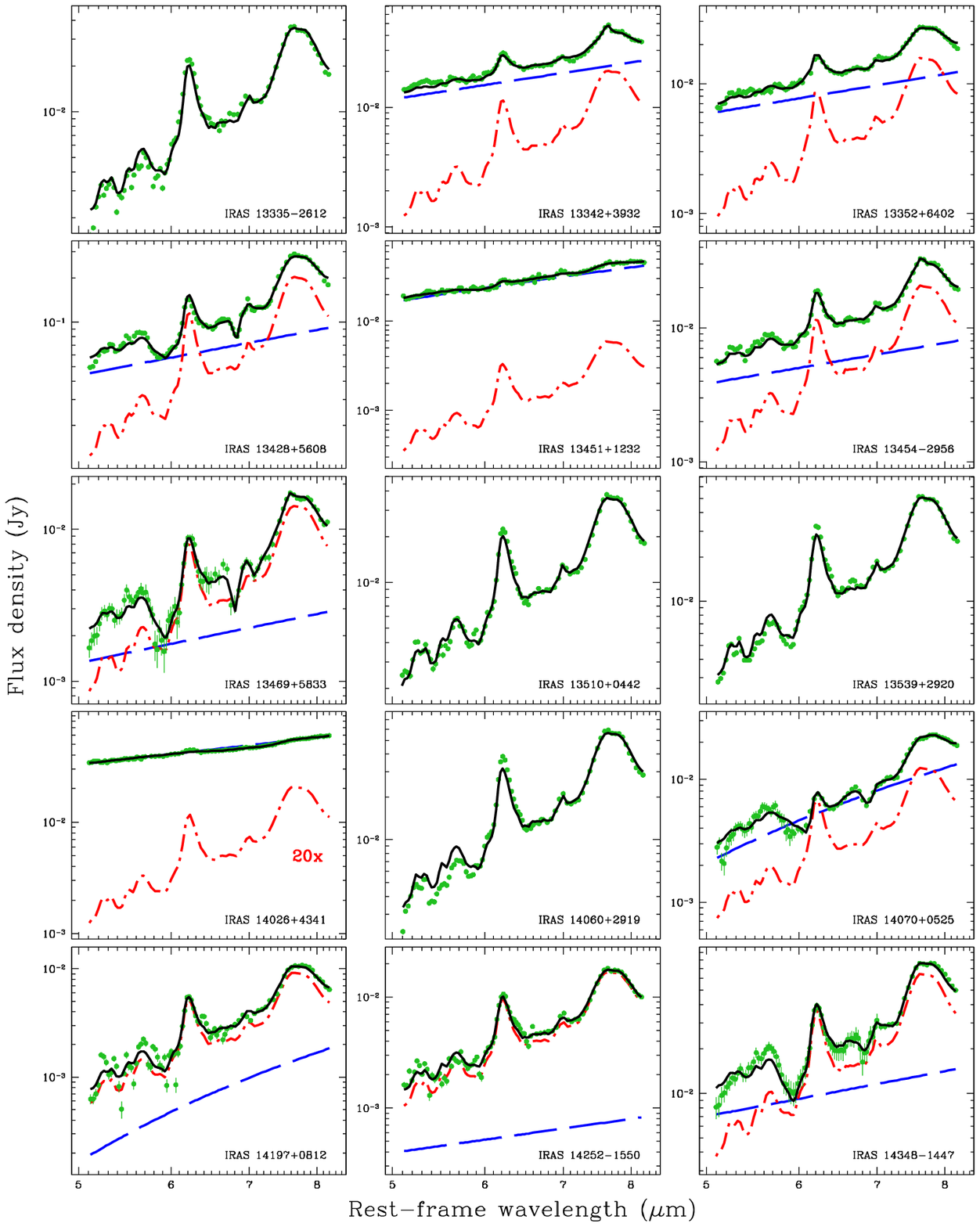}
\contcaption{} 
\end{figure*}
\begin{figure*}
\includegraphics[width=\linewidth]{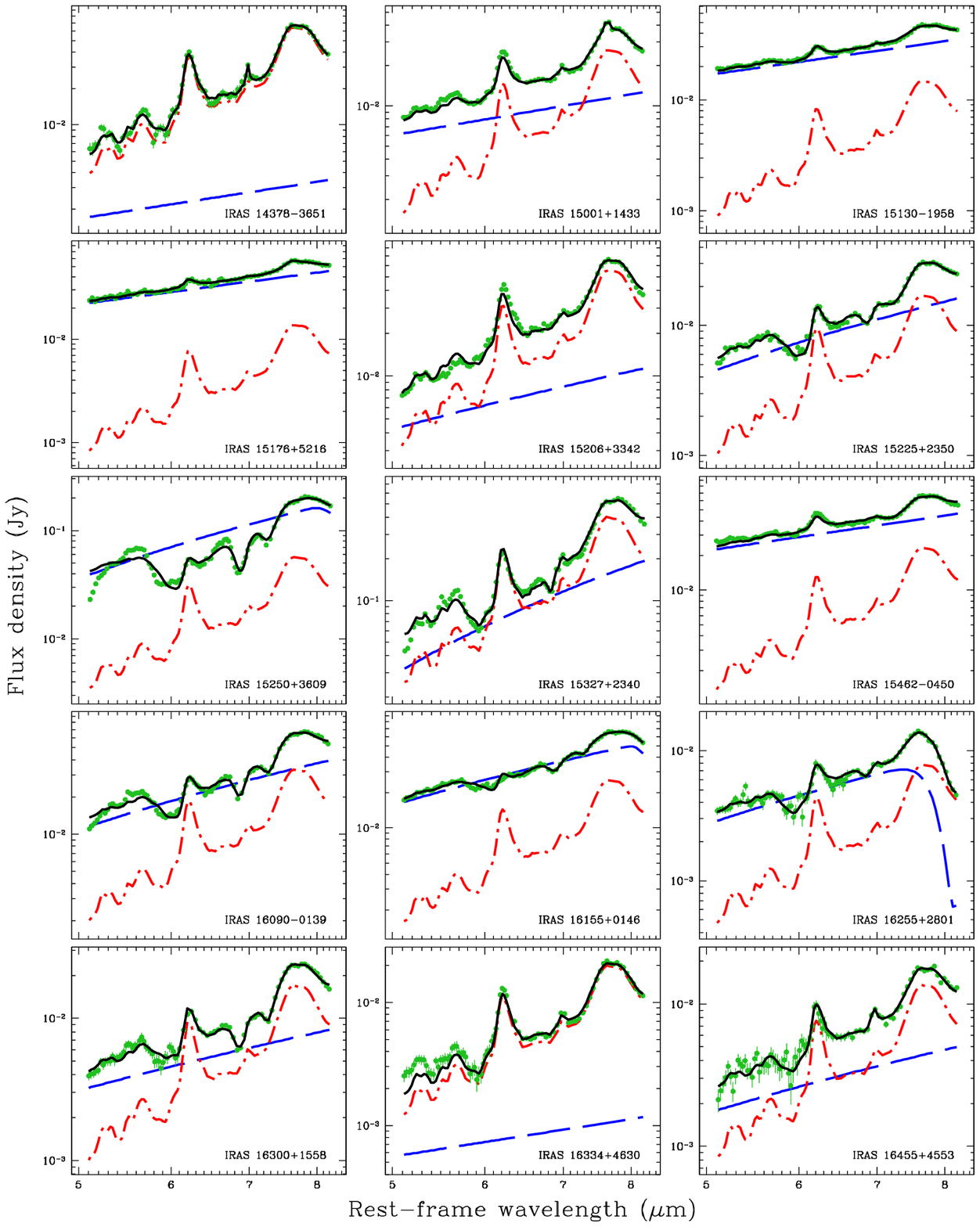}
\contcaption{} 
\end{figure*}
\begin{figure*}
\includegraphics[width=\linewidth]{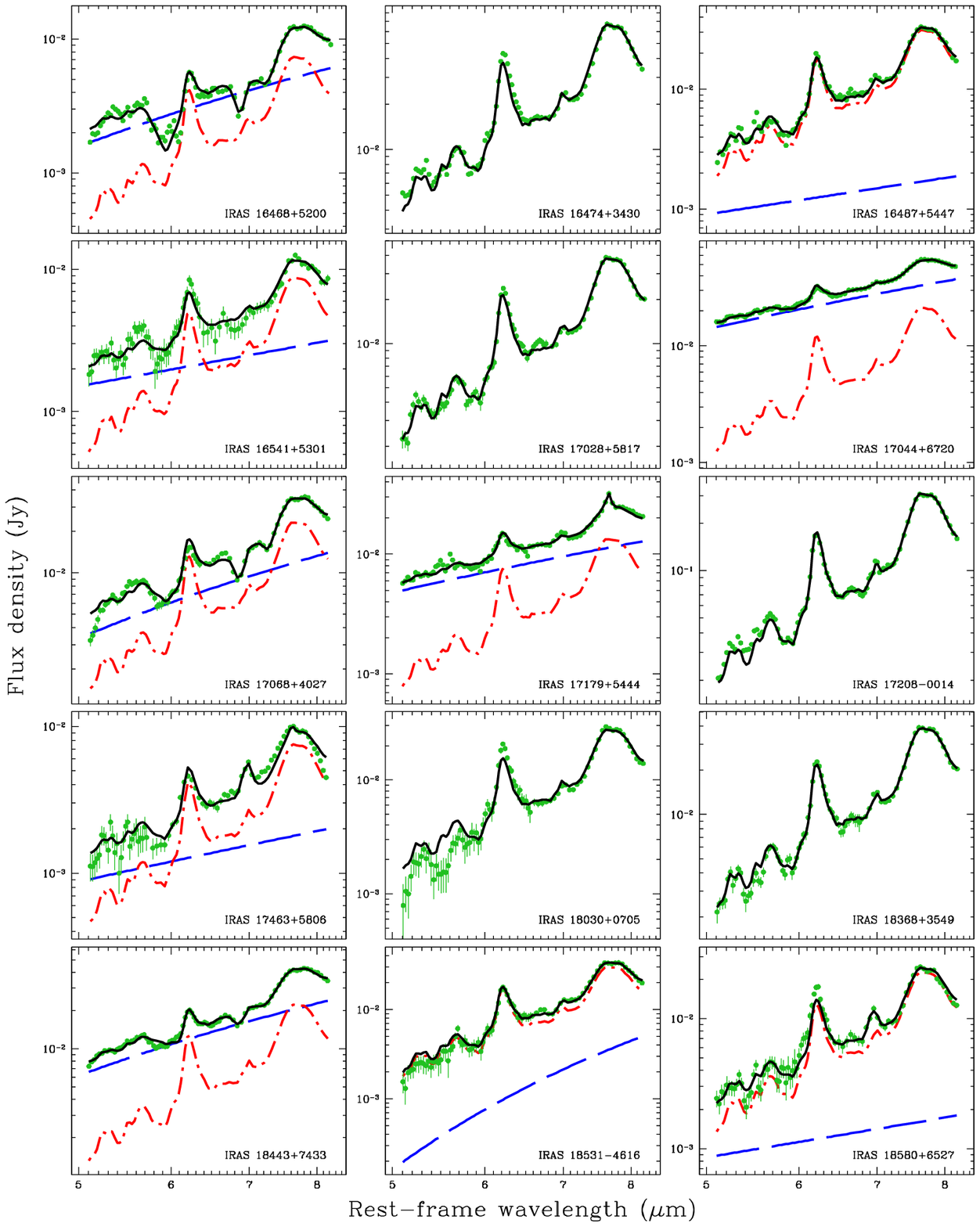}
\contcaption{} 
\end{figure*}
\begin{figure*}
\includegraphics[width=\linewidth]{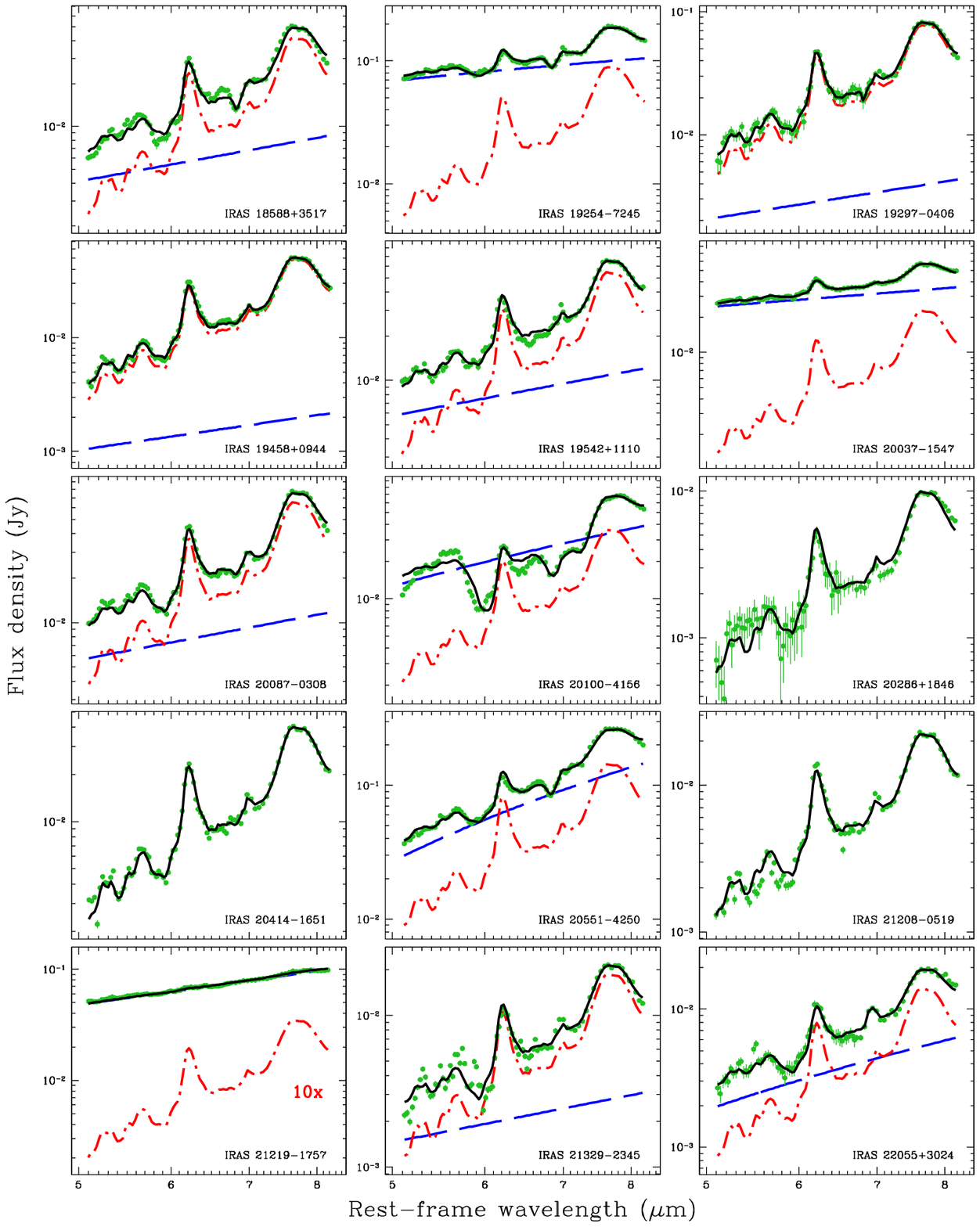}
\contcaption{} 
\end{figure*}
\begin{figure*}
\includegraphics[width=\linewidth]{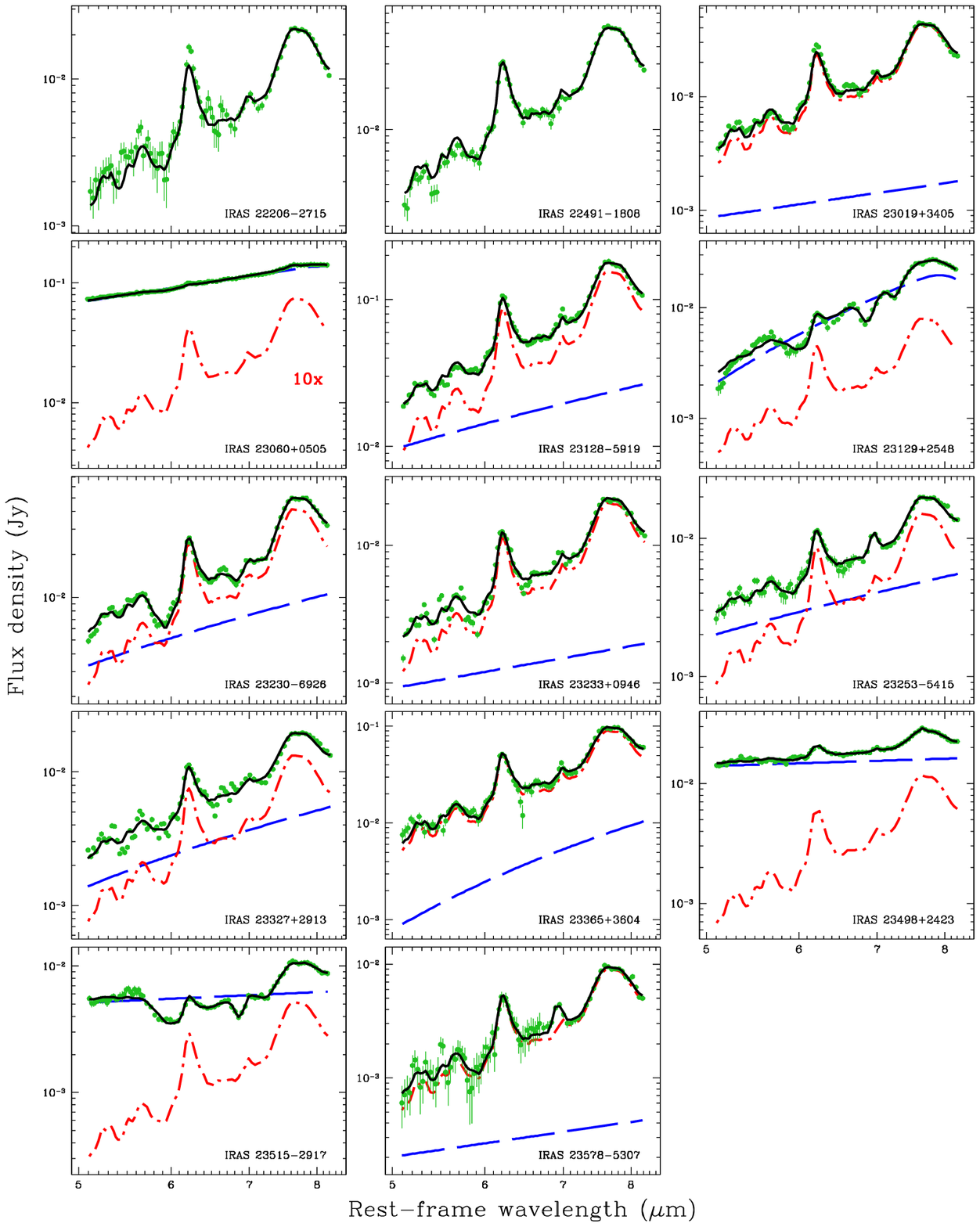}
\contcaption{} 
\end{figure*}

\label{lastpage}

\end{document}